\documentclass{article}
\usepackage{graphicx}
\usepackage{amsmath}
\usepackage{multirow}
\usepackage{amssymb}
\usepackage{booktabs}
\usepackage{authblk}
\usepackage{caption}

\usepackage[letterpaper,margin=1in]{geometry}
\usepackage[backend=biber, style = authoryear, maxcitenames = 1, uniquelist=false, uniquename=false]{biblatex}
\usepackage[colorlinks=true,linkcolor=blue,citecolor=blue,urlcolor=blue]{hyperref}
\usepackage[nameinlink]{cleveref}
\usepackage{float}

\DeclareFieldFormat{citehyperref}{%
  \DeclareFieldAlias{bibhyperref}{noformat}%
  \bibhyperref{#1}%
}
\savebibmacro{cite}
\renewbibmacro*{cite}{%
  \printtext[citehyperref]{%
    \restorebibmacro{cite}%
    \usebibmacro{cite}}}

\title{Cross-modal dependence analysis with asynchronous longitudinal multimodal data}
\author{KUN QIAN, HYUNG G. PARK$^\ast$ \\
\textit{Division of Biostatistics, Department of Population Health, New York University Grossman School of Medicine, 180 Madison Ave., New York, NY 10016, USA}\\
\vspace{0.5em}
FOR THE ALZHEIMER'S DISEASE NEUROIMAGING INITIATIVE $^\dagger$\\
\vspace{0.5em}
\small *Corresponding author. Email: parkh15@nyu.edu\\
$^\dagger$ Data used in preparation of this article were obtained from the Alzheimer’s Disease Neuroimaging Initiative (ADNI)
database (\url{adni.loni.usc.edu}). As such, the investigators within ADNI contributed to the design and implementation of
ADNI and/or provided data but did not participate in analysis or writing of this report. A complete listing of ADNI
investigators can be found at:  \url{http://adni.loni.usc.edu/wpcontent/uploads/how_to_apply/ADNI_Acknowledgement_List.pdf}\\
}
\date{June 26, 2026}

\begin{document}

\maketitle
\section*{SUMMARY} 
We propose a Bayesian latent variable model to characterize covariate-specific dependence structures among multiple modalities of asynchronously collected multivariate data. This setting commonly arises in longitudinal biomedical research, especially in observational and clinical studies of complex diseases, where dynamic and heterogeneous dependence across biomarker modalities can be biologically and clinically informative. However, quantitative analysis is often challenged by asynchronous collection of multimodal profiles due to study design and data collection constraints. For example, the biological diagnosis and staging of Alzheimer's disease require integrated evaluation of multimodal biomarkers, including imaging and biofluid biomarkers, and the Alzheimer's Disease Neuroimaging Initiative (ADNI) study has collected biomarker profiles longitudinally on varying schedules for over two decades. Common analytic strategies that rely solely on complete multimodal profiles or analyze each modality separately can result in information loss and biased estimates. Therefore, we aim to jointly incorporate all available observations to estimate the population-level cross-modal dependence structures (e.g., covariance or correlation matrices) that evolve over time and vary across demographic or clinical groups. The proposed model uses modality-specific low-rank loading matrices with shared latent variables to integrate information across modalities, visits, and subjects, while accounting for repeated measurements. The application to ADNI data reveals clinically meaningful patterns in longitudinal cross-modal biomarker dependence, and the simulation study shows improved recovery under limited modality synchrony. \\ 
\\
\textbf{Keywords:} asynchronous multimodal data, Bayesian inference, covariance regression, latent variable model, longitudinal data analysis, multivariate data, observational studies
\section{INTRODUCTION}
Understanding the dynamics and heterogeneity in dependence across multiple modalities of biomarkers facilitates biological, pathophysiological, and therapeutic research (\cite{strimbu2010}, \cite{califf2018}), especially for complex diseases that are chronic and influenced by various factors (\cite{schork1997}, \cite{johansson2023}). For example, the biological diagnosis and staging of Alzheimer's disease (AD) require joint consideration of multimodal biomarker profiles, such as amyloid and tau positron emission tomography (PET), cerebrospinal fluid (CSF), and plasma biomarkers, and pathologies may vary across populations (\cite{therriault2024}, \cite{jack2024}). Quantitatively, the data profiles of many biomarker modalities can be naturally represented as multivariate random vectors, and the joint variability across modalities can be summarized using the cross-covariance matrix, which has served as a classical measure of dependence between two random vectors and underlies many statistical methods, such as canonical correlation analysis (\cite{andrew2013}, \cite{leon2017}).

\begin{figure}[htbp]
    \centering
    \includegraphics[width=0.7\textwidth]{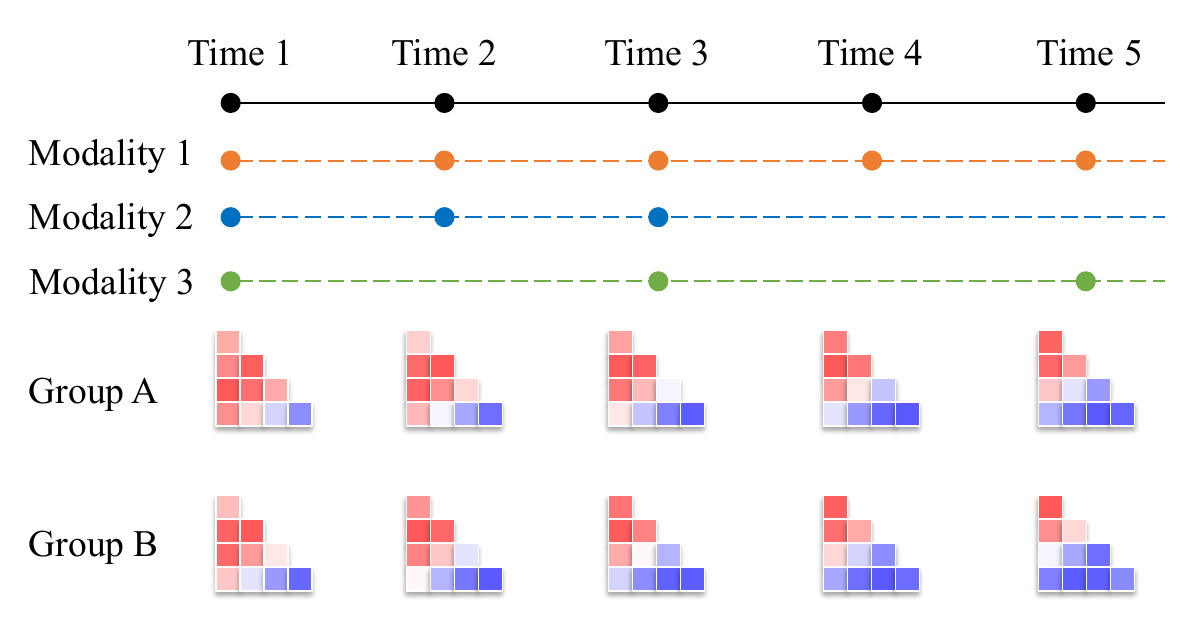}
    \caption{
        Illustration of the motivating study design (top panel) and model outputs (bottom panel). The multimodal data may be collected asynchronously such that modality profiles are only partially obtained at some time points. The proposed model leverages these data to characterize time-varying and subgroup-specific within- and between-modality dependence structures, such as covariance or correlation matrices.
    }
    \label{fig:illustration}
\end{figure}

The increasing prevalence of observational studies and clinical trials that longitudinally collect multiple biomarker profiles has generated rich datasets for quantitative studies of disease pathophysiology (\cite{abe2022}, \cite{edwards2023}, \cite{he2025longitudinal}, \cite{trieu2025longitudinal}). In practice, different modalities are usually collected asynchronously over the study period (\Cref{fig:illustration}) such that some data records contain measurements for all modalities, whereas others contain only a subset (\Cref{tab:data_structure}). For example, the Alzheimer's Disease Neuroimaging Initiative (ADNI) has collected longitudinal multimodal data over more than two decades, including clinical evaluations, neuroimaging measures (e.g., magnetic resonance imaging [MRI] and PET), fluid biomarkers (e.g., CSF and plasma), genetic data, and neuropsychological assessments. These data were obtained according to modality-specific sampling schedules, resulting in a rich but highly asynchronous longitudinal dataset (\cite{petersen2010}, \cite{veitch2024alzheimer}, \cite{weiner2025overview}, \cite{kanoria2026clinical}). In addition, the Integrative Human Microbiome Project (HMP2 or iHMP) generated clinical and multi-omics profiles of host and microbial activity from biopsy, blood, and stool samples collected on varying schedules (e.g., weekly, biweekly, and quarterly) to investigate inflammatory bowel diseases (\cite{lloyd2019multi}). Nevertheless, there remains a need for targeted methodological tools that can fully leverage the complexity of longitudinal asynchronous multimodal data.

\begin{table}[ht]
\centering
\caption{Example of Asynchronous Multimodal Longitudinal Data Structure}
\label{tab:data_structure}
\begin{tabular}{ccccccc}
\toprule
\multirow{3}{*}{ID}
& \multicolumn{3}{c}{Responses}
& \multicolumn{3}{c}{Covariates} \\
\cmidrule(lr){2-4}
\cmidrule(lr){5-7}
& Modality 1
& Modality 2
& Modality 3
& \multirow{2}{*}{Time}
& \multirow{2}{*}{Group}
& \multirow{2}{*}{$\cdots$} \\
& (5 variables)
& (10 variables)
& (7 variables)
& &
& \\
\midrule
1 & $\times$ & $\times$ & $\times$ & 1 & A & \\
1 &           & $\times$ & $\times$ & 3 & A & \\
1 & $\times$ &           & $\times$ & 5 & A & $\cdots$ \\
2 & $\times$ & $\times$ & $\times$ & 1 & B & \\
2 & $\times$ &           &          & 2 & B & \\
\bottomrule
\end{tabular}

\vspace{0.5em}

\begin{minipage}{0.75\linewidth}
\footnotesize
Note: $\times$ indicates that all variables from the corresponding modality are collected at the visit.
\end{minipage}
\end{table}

Existing methods for asynchronous longitudinal data typically focus on settings in which responses and covariates are observed irregularly on different time grids, using pre-processing (\cite{xiong2010binning}), kernel-weighted estimation (\cite{csenturk2013modeling}, \cite{cao2015regression}, \cite{cao2016last}, \cite{chen2017analysis}, \cite{sun2021regression}, \cite{liu2023regression}, \cite{zhong2024locally}) and functional data analysis techniques (\cite{chang2023asynchronous}, \cite{lu2025semiparametric}) for inference. Recent developments have further enabled the incorporation of functional covariate processes (\cite{li2022regression}, \cite{li2023asynchronous}). The objective of these methods is to quantify the time-invariant or time-varying association between two longitudinal processes. Additionally, \cite{liu2026comprehensive} proposed to estimate bivariate time-varying odds ratio and relative risk to capture the dynamic relationships between two binary variables. In contrast to existing approaches, the proposed work is developed for a different scenario. Here, asynchronous multimodal data refer to settings in which different modalities of multivariate response vectors may not be simultaneously collected, so that each record contains all or only a subset of the modalities (\Cref{tab:data_structure}). The covariates are either time-invariant or observed concurrently with the available response measurements. Rather than to model the association between a response process and a covariate process, our objective is to estimate the dependence structures across multiple modalities of responses conditional on covariates. Credible intervals and other posterior summaries can be reported for each element of the derived covariance matrix with the aim of quantify uncertainty in pairwise dependence. The proposed model is suitable for observational and clinical studies with a limited number of repeated measurements per subject, making subject-specific smoothing or trajectory estimation unreliable for standard functional data analysis methods. Existing practice relies primarily on empirical procedures that restrict analysis to complete cases with jointly observed modalities within covariate-based strata, leading to loss of information and potential bias in parameter estimation, especially when the total number of observations is limited.

We therefore propose a latent variable model to extract leading covariate-related signals from noisy real-world data. For illustrative purposes, when the study involves a total of $K$ modalities of multivariate data, let $\boldsymbol{y}^{(k)}_{ij}$ denote the $k^{th}$ modality for subject $i$ at visit $j$, and $\boldsymbol{y}_{ij}=(\boldsymbol{y}_{ij}^{(1)T}, \cdots, \boldsymbol{y}_{ij}^{(K)T})^T$ denote the complete data vector stacking all modalities. A standard latent variable model can be expressed as
\[\boldsymbol{y}_{ij} = \boldsymbol{\Gamma}\boldsymbol{\eta}_{ij} + \boldsymbol{\epsilon}_{ij}\] 
where $\boldsymbol{\Gamma}$ represents a low-rank dimension reduction matrix across all modalities, and $\boldsymbol{\eta}_{ij}$ represents the latent variable vector with fewer dimensions than the data vector $\boldsymbol{y}_{ij}$, and $\boldsymbol{\epsilon}_{ij}$ represents the truncated components. When the multimodal data are asynchronously observed such that only a subvector of the concatenated vector $\boldsymbol{y}_{ij}$ is collected, we define $\boldsymbol{\Gamma}=(\boldsymbol{\Gamma}^{(1)T},\cdots, \boldsymbol{\Gamma}^{(K)T})^T$, and for each observation, $\boldsymbol{\eta}_{ij}$ is shared across modalities and only the blocks of $\boldsymbol{\Gamma}$ corresponding to the observed subvector may contribute to the likelihood. 
For instance, when three modalities are analyzed, if only modalities 1 and 3 were observed for a specific observation such that  $\boldsymbol{y}_{ij}^{obs}=(\boldsymbol{y}_{ij}^{(1)T}, \boldsymbol{y}_{ij}^{(3)T})^T$, then the likelihood for this data point would involve only $(\boldsymbol{\Gamma}^{(1)T}, \boldsymbol{\Gamma}^{(3)T})^T$. For the latent variable $\boldsymbol{\eta}_{ij}$, we adopt a mixed-effect formulation 
\[\boldsymbol{\eta}_{ij} = \boldsymbol{\mu}(t_{ij}, \boldsymbol{x}_{ij}) + \boldsymbol{b}_i + \boldsymbol{u}_{ij}, \quad
\boldsymbol{u}_{ij}\sim \mathcal{N}\!\left(0,\boldsymbol{\Omega}(t_{ij}, \boldsymbol{x}_{ij})\right)\]
consisting of fixed effects $\boldsymbol{\mu}(t_{ij},\boldsymbol{x}_{ij})$, subject-level random effects $\boldsymbol{b}_i$ and residual components $\boldsymbol{u}_{ij}$, where both fixed effects and residuals are associated with covariates (e.g., time $t_{ij}$ and demographic or clinical variables $\boldsymbol{x}_{ij}$). We employ a covariance regression approach (\cite{park2025bayesian}) to estimate the time- and subgroup-dependent residual covariance $\boldsymbol{\Omega}(t_{ij},\boldsymbol{x}_{ij})$ in the shared latent space.

The primary outputs are cross-covariance matrices $\boldsymbol{\Gamma}^{(k)}\boldsymbol{\Omega}(t_{ij},\boldsymbol{x}_{ij})\boldsymbol{\Gamma}^{(k')T}$ for any two modalities $k$ and $k'$, which characterize the time- and subgroup-specific dependence structures. They can be normalized into cross-correlation matrices for scale-free interpretation.
Notably, the model outputs are interpreted at the population level rather than the individual level. Although modeling covariate-dependent random effects (e.g., allowing $\boldsymbol{b}_i$ to vary with $t_{ij}$ and $\boldsymbol{x}_{ij}$) provides more direct characterization of subject-level mechanisms, it usually requires a large number of repeated measurements for accurate inference (\cite{laird1982random}, \cite{backenroth2018modeling}). In biomedical observational or clinical studies, the number of repeated observations per subject is typically limited for biomarker data. Therefore, the proposed analysis focuses on population-level rather than individualized dependence structure. Subject-specific random effects are introduced to account for dependence induced by repeated measurements, serving as a nuisance adjustment as is typical in applied mixed-effect models. Further, this paper incorporates asynchronous multivariate responses from longitudinal studies to estimate and interpret covariate-dependent residual covariance, addressing an inferential objective distinct from longitudinal modeling of covariance matrix-valued outcomes (\cite{zhao2024longitudinal}). 

In summary, the goal of this work is to fully utilize real-world longitudinal multimodal data to characterize population-level dependence across biomarker modalities over time and across heterogeneous groups, thereby providing quantitative summaries for biological and clinical evaluation. Statistically, the proposed method aims (i)
to model covariate-dependent residual covariance matrices for Gaussian random vectors jointly with fixed and random effects;
(ii) to incorporate asynchronously observed random vectors through a multivariate statistical approach. This paper is organized as follows: \Cref{sec:methods} describes the model formulation and inferential procedures. In \Cref{sec:simulation}, we conduct simulation studies to evaluate model performance under various data scenarios and investigate the impact of key model parameters. \Cref{sec:application} illustrates the proposed method through application to ADNI data. Concluding remarks are provided in \Cref{sec:discussion}. 

\section{METHODS}\label{sec:methods}
\subsection{Notations and overview}\label{sec:notation}
In a study investigating $K$ modalities of multivariate data, let $\boldsymbol{y}^{(k)}_{ij}=(y^{(k)}_{ij1}, \cdots, y^{(k)}_{ijp_k})^T\in \mathbb{R}^{p_k}$ denote the $k^{th}$ ($k=1,\cdots, K$) modality for subject $i$ $(i=1,\cdots,N)$ at visit $j$ ($j=1,\cdots,n_i$), and when all modalities are simultaneously collected, the complete data vector is $\boldsymbol{y}_{ij}=(\boldsymbol{y}_{ij}^{(1)T}, \cdots, \boldsymbol{y}_{ij}^{(K)T})^T=(y_{ij1}^{(1)},\cdots,y_{ijp_1}^{(1)},\cdots,y_{ij1}^{(K)},\cdots,y_{ijp_K}^{(K)})^T\in \mathbb{R}^p$, where $p=\sum_{k=1}^K p_k$. For this scenario, the standard low-rank latent variable model $\boldsymbol{y}_{ij} = \boldsymbol{\Gamma}\boldsymbol{\eta}_{ij} + \boldsymbol{\epsilon}_{ij}$ as described in the introduction section is parameterized as follows: $\boldsymbol{\Gamma}\in \mathbb{R}^{p\times d}$ has orthogonal columns with $d< p$, and $\boldsymbol{\eta}_{ij}\in \mathbb{R}^d$ represents the latent variable of interest, and $\boldsymbol{\epsilon}_{ij}\in \mathbb{R}^{p}$ denotes the truncated nuisance components that lie in the orthogonal complement of the column space spanned by $\boldsymbol{\Gamma}$.

In practice, only a subset of modalities may be observed for each data point. Let $\mathcal{M}_{ij}$ indicate the index set of observed modalities for subject $i$ at visit $j$, and $m_{ij}=|\mathcal{M}_{ij}|$ denote the number of modalities observed, where $m_{ij}\in \{1,\cdots, K\}$. For instance, when modeling 3 modalities in a study, $\mathcal{M}_{ij}=\{1,3\}$ represents that the 1st and 3rd modalities are observed for visit $j$ from subject $i$, and in this case, $m_{ij}=2$. Then the observed data $\boldsymbol{y}_{ij}^{obs}\in \mathbb{R}^{p_{ij}}$ is a stack of $\boldsymbol{y}_{ij}^{(k)}$ for all $k \in \mathcal{M}_{ij}$ with $p_{ij}={\sum_{k\in \mathcal{M}_{ij}} p_k}$. To accommodate this common practical complexity, we use modality-specific orthonormal basis matrices $\boldsymbol{\Gamma}^{(k)}\in \mathbb{R}^{p_k\times d}$ with $d\le \min_k p_k$ to form $\boldsymbol{\Gamma}=(\boldsymbol{\Gamma}^{(1)T},\cdots, \boldsymbol{\Gamma}^{(K)T})^T$ satisfying $\boldsymbol{\Gamma}^T\boldsymbol{\Gamma}=K\mathbf{I}_d$.

Additionally, the proposed model can incorporate covariate adjustments. Throughout, we let $t_{ij}$ denote the visit time, and $\boldsymbol{x}_{ij}\in \mathbb{R}^{q_0}$ denote a vector of time-fixed or time-varying demographic or clinical variables.

\subsection{Model formulation}
We propose a Bayesian latent variable approach to estimate the population-level cross-covariance matrices at specific time points and subgroups, while incorporating asynchronous multimodal data. The modality $k$-specific formulation is
\begin{equation}\label{eq:main-modality}
\boldsymbol{y}_{ij}^{(k)} = \boldsymbol{\Gamma}^{(k)}\boldsymbol{\eta}_{ij} + \boldsymbol{\epsilon}_{ij}^{(k)}, \quad 
\boldsymbol{\eta}_{ij} = \boldsymbol{\mu}(t_{ij}, \boldsymbol{x}_{ij}) + \boldsymbol{b}_i + \boldsymbol{u}_{ij}, \quad
\boldsymbol{u}_{ij}\sim \mathcal{N}\!\left(\boldsymbol{0},\boldsymbol{\Omega}(t_{ij}, \boldsymbol{x}_{ij})\right)
\end{equation}
where $\boldsymbol{\Gamma}^{(k)}$ constitutes a subset of the eigenbasis of $\mathrm{Var}(\boldsymbol{y}_{ij}^{(k)})$ that facilitates model inference through the likelihood derivation such that $d\le p_k$ across all $k$, and $\boldsymbol{\epsilon}^{(k)}_{ij}\in \mathbb{R}^{p_k}$ represents the truncated nuisance part that lies in the orthogonal complement of the column space spanned by  $\boldsymbol{\Gamma}^{(k)}$. Moreover, we decompose the latent variable $\boldsymbol{\eta}_{ij}\in \mathbb{R}^d$ into fixed effects $\boldsymbol{\mu}(t_{ij}, \boldsymbol{x}_{ij})$, random effects $\boldsymbol{b}_i$, and residual $\boldsymbol{u}_{ij}$. We assume that $\boldsymbol{b}_i\sim \mathcal{N}(\boldsymbol{0},\boldsymbol{\Omega}_{\boldsymbol{b}})$ where $\boldsymbol{\Omega}_{\boldsymbol{b}}\in \mathrm{Sym}^+_d$ is diagonal and $\boldsymbol{\Omega}(t_{ij}, \boldsymbol{x}_{ij})\in \mathrm{Sym}^+_d$ models the latent core residual covariance. The target conditional cross-covariance for any pair of modalities (e.g., modalities $k$ and $k'$) is
$$ 
\mathrm{Cov}(\boldsymbol{y}^{(k)}_{ij},\boldsymbol{y}^{(k')}_{ij}|\boldsymbol{b}_i)=\boldsymbol{\Gamma}^{(k)}\boldsymbol{\Omega}(t_{ij},\boldsymbol{x}_{ij})\boldsymbol{\Gamma}^{(k')T}.
$$
Furthermore, the fixed effects $\boldsymbol{\mu}(t_{ij}, \boldsymbol{x}_{ij})$ can be evaluated while accounting for heteroskedasticity. Other quantities, such as within-modality covariance (e.g., $\boldsymbol{\Gamma}^{(k)}\boldsymbol{\Omega}(t_{ij},\boldsymbol{x}_{ij})\boldsymbol{\Gamma}^{(k)T}$), can also be assessed, and correlations can be subsequently derived. Notably, these characteristics are interpreted at the population level while adjusting for individual heterogeneity. 

\subsection{Parameter specification}
We further define the observation-specific parametrization of model \eqref{eq:main-modality} to illustrate the inference process, because each observation $\boldsymbol{y}_{ij}^{obs}$ as specified in \Cref{sec:notation} contains all or part of the modalities (e.g., $p_{ij}$ may not equal $p_{ij'}$ for subject $i$ at visits $j$ and $j'$) and the estimation of the latent variable $\boldsymbol{\eta}_{ij}$ jointly involves all observed modalities at each observation. Let
\begin{equation}\label{eq:main-obs}
\boldsymbol{y}_{ij}^{obs} = \boldsymbol{\Gamma}_{ij}\boldsymbol{\eta}_{ij} + \mathbf{L}_{ij}\boldsymbol{e}_{ij}
\end{equation}
where $\boldsymbol{\Gamma}_{ij}\in \mathbb{R}^{p_{ij}\times d}$ denotes the dimension reduction matrix corresponding to $\boldsymbol{y}_{ij}^{obs}$ designed as the vertical concatenation of $\boldsymbol{\Gamma}^{(k)}$ for all $k \in \mathcal{M}_{ij}$ with $\boldsymbol{\Gamma}_{ij}^{T}\boldsymbol{\Gamma}_{ij}=m_{ij}\mathbf{I}_d$, and therefore $m_{ij}^{-1/2}\boldsymbol{\Gamma}_{ij}$ has orthonormal columns, and $\mathbf{L}_{ij}\in \mathbb{R}^{p_{ij}\times (p_{ij}-d)}$ denotes the truncated complement directions with $\boldsymbol{\Gamma}^T_{ij}\mathbf{L}_{ij}=\mathbf{0}$. Then $\begin{bmatrix}m_{ij}^{-1/2}\boldsymbol{\Gamma}_{ij} & \mathbf{L}_{ij}(\mathbf{L}_{ij}^T\mathbf{L}_{ij})^{-1/2}\end{bmatrix}$ forms an orthonormal eigenbasis of $\boldsymbol{\Sigma}_{ij}=\mathrm{Var}(\boldsymbol{y}_{ij}^{obs})$. $\boldsymbol{e}_{ij}\in \mathbb{R}^{p_{ij}-d}$ collects the nuisance components in the truncated directions.

For simplified illustration, let $\tilde{\boldsymbol{x}}_{ij}=(1,t_{ij}, \boldsymbol{x}^T_{ij})^T\in \mathbb{R}^q, q=q_0+2$ denote the design vector for all covariates. The fixed effects and residual covariance in the latent core $\boldsymbol{\eta}_{ij}$ as in model \eqref{eq:main-modality} are represented as
\[
\boldsymbol{\mu}(\tilde{\boldsymbol{x}}_{ij})=\mathbf{A} \tilde{\boldsymbol{x}}_{ij}, \quad \boldsymbol{\Omega}(\tilde{\boldsymbol{x}}_{ij})=\mathrm{diag}(\exp(\mathbf{B} \tilde{\boldsymbol{x}}_{ij}))
\]
where $\mathbf{A}, \mathbf{B}\in \mathbb{R}^{d\times q}$ are coefficient matrices and $\mathrm{diag}(\cdot)$ denotes the operator that constructs a diagonal matrix from a vector. Unlike standard regression settings where both responses and covariates lie in vector spaces, $\boldsymbol{\Omega}(\tilde{\boldsymbol{x}}_{ij})$ is a symmetric positive definite matrix in a curved manifold. The manifold geometry complicates direct modeling, and we adopt the matrix whitening transport that maps the covariance matrices into a common tangent space, enabling regression modeling via linear mappings between vector spaces (\cite{ng2015transport}, \cite{park2025bayesian}). In detail, the mapping is given by
\begin{equation}\label{eq:whitening}
\phi(\boldsymbol{\Omega}(\tilde{\boldsymbol{x}}_{ij}))=\log(\bar{\boldsymbol{\Omega}}^{-1/2}_{ij}\boldsymbol{\Omega}(\tilde{\boldsymbol{x}}_{ij})\bar{\boldsymbol{\Omega}}^{-1/2}_{ij})=\mathrm{diag}(\tilde{\mathbf{B}}\tilde{\boldsymbol{x}}_{ij})
\end{equation}
where $\bar{\boldsymbol{\Omega}}^{-1/2}_{ij}$ is the whitening matrix that brings $\boldsymbol{\Omega}(\tilde{\boldsymbol{x}}_{ij})$ close to $\mathbf{I}_d$, which serves as the reference point for tangent space representations obtained through the log map, and $\tilde{\mathbf{B}}\in \mathbb{R}^{d\times q}$ is the directly inferred coefficient matrix in the tangent space. The correspondence between the estimate of $\tilde{\mathbf{B}}$ and $\mathbf{B}$ can be analytically derived (Section S2 in Supplementary Material). Particularly, the whitening matrix is designed as $\bar{\boldsymbol{\Omega}}^{-1/2}_{ij}:=\boldsymbol{\Gamma}_{ij}^{T}\bar{\boldsymbol{\Sigma}}^{-1/2}_{ij}\boldsymbol{\Gamma}_{ij}$ with $\bar{\boldsymbol{\Sigma}}_{ij}\in\mathrm{Sym}_{p_{ij}}^+$ indicating the sample mean covariance across modalities $k\in \mathcal{M}_{ij}$, which acts as a practical approximation that can be estimated from the observed data as it also contains variation in random effects. 

\subsubsection{Identifiability}
We constrain $\mathbf{A}$ to be upper triangular with positive diagonal entries to ensure identifiability for computational stability, following the convention of the thin QR decomposition (\cite{golub2013matrix}). The parameterization in the fixed effect involves the product $\boldsymbol{\Gamma}_{ij}\mathbf{A}$, which remains the same for any nonsingular rotation applied on $\boldsymbol{\Gamma}_{ij}$ and $\mathbf{A}$. Although the individual matrix is not of interest for interpretation and an identifiable combination of the quantities is sufficient for estimation, the ambiguity causes practical problems in Bayesian sampling, such as slow mixing and multiple equivalent modes due to equivalent representation of the same likelihood.

\subsubsection{Rank determination}
The proposed model employs a truncated estimation approach to recover the main underlying structure, and the rank $d$ is designed as a pre-defined parameter in this model. We conducted simulation studies to examine its impact and to guide practical use.

\subsection{Bayesian inference}
Inference focuses exclusively on the subspace spanned by $\boldsymbol{\Gamma}_{ij}$ in equation \eqref{eq:main-obs}, whereas the subspace spanned by $\mathbf{L}_{ij}$ and its associated parameters are treated as nuisance components and are not estimated. Using $\boldsymbol{\Gamma}^T_{ij} \mathbf{L}_{ij}=\mathbf{0}$, the likelihood for the modeled subspace reduces to
\[
p(\boldsymbol{y}_{ij}^{obs}|\boldsymbol{\Gamma}_{ij},\mathbf{A},\mathbf{\tilde{B}},\boldsymbol{b}_i,\boldsymbol{\Omega}_{\boldsymbol{b}},\tilde{\boldsymbol{x}}_{ij})\propto
|\boldsymbol{\Omega}(\tilde{\boldsymbol{x}}_{ij})|^{-1/2}
\] 
\[
\exp\left(-\frac12 (\boldsymbol{y}_{ij}^{obs}-\boldsymbol{\Gamma}_{ij}\mathbf{A}\tilde{\boldsymbol{x}}_{ij}-\boldsymbol{\Gamma}_{ij}\boldsymbol{b}_i)^T\boldsymbol{\Gamma}_{ij}\boldsymbol{\Omega}^{-1}(\tilde{\boldsymbol{x}}_{ij})\boldsymbol{\Gamma}_{ij}^T (\boldsymbol{y}_{ij}^{obs}-\boldsymbol{\Gamma}_{ij}\mathbf{A}\tilde{\boldsymbol{x}}_{ij}-\boldsymbol{\Gamma}_{ij}\boldsymbol{b}_i)\right)
\]
The detailed derivation is provided in Section S1 of the Supplementary Material. Under the working spectral parameterization that $m_{ij}^{-1/2}\boldsymbol{\Gamma}_{ij}$ is a subset of the eigenbasis of $\boldsymbol{\Sigma}_{ij}$, let $\boldsymbol{\bar{\Lambda}}_{ij}\in \mathbb{R}^{d\times d}$ denote the diagonal matrix whose diagonal entries are the eigenvalues of $\bar{\boldsymbol{\Sigma}}_{ij}$ corresponding to $m_{ij}^{-1/2}\boldsymbol{\Gamma}_{ij}$, and therefore $\bar{\boldsymbol{\Omega}}^{1/2}_{ij}=\boldsymbol{\Gamma}_{ij}^{T}\bar{\boldsymbol{\Sigma}}^{1/2}_{ij}\boldsymbol{\Gamma}_{ij}=m_{ij}\boldsymbol{\bar{\Lambda}}_{ij}^{1/2}$, which does not depend on any model parameters. By equation \eqref{eq:whitening}, the determinant term can be expressed as
\[
|\boldsymbol{\Omega}(\tilde{\boldsymbol{x}}_{ij})|=|\bar{\boldsymbol{\Omega}}^{1/2}_{ij}\mathrm{diag}(\exp(\tilde{\mathbf{B}}\tilde{\boldsymbol{x}}_{ij}))\bar{\boldsymbol{\Omega}}^{1/2}_{ij}|
\propto |\mathrm{diag}(\exp(\tilde{\mathbf{B}}\tilde{\boldsymbol{x}}_{ij}))|
\]
For the covariance components in the exponentiation, since
$\boldsymbol{\Gamma}_{ij}\bar{\boldsymbol{\Omega}}^{-1/2}_{ij}=m^{-1}_{ij}\boldsymbol{\Gamma}_{ij}\bar{\boldsymbol{\Lambda}}_{ij}^{-1/2}=m^{-1}_{ij}\bar{\boldsymbol{\Sigma}}^{-1/2}_{ij}\boldsymbol{\Gamma}_{ij}$, then
\begin{align*}
\boldsymbol{\Gamma}_{ij}\boldsymbol{\Omega}^{-1}(\tilde{\boldsymbol{x}}_{ij})\boldsymbol{\Gamma}_{ij}^{T} 
=&\boldsymbol{\Gamma}_{ij}\bar{\boldsymbol{\Omega}}^{-1/2}_{ij}\{\mathrm{diag}(\exp(\tilde{\mathbf{B}}\tilde{\boldsymbol{x}}_{ij}))\}^{-1}\bar{\boldsymbol{\Omega}}^{-1/2}_{ij}\boldsymbol{\Gamma}_{ij}^{T}
\\
=& (m^{-1}_{ij}\boldsymbol{\Gamma}_{ij}^T\bar{\boldsymbol{\Sigma}}^{-1/2}_{ij})^T\{\mathrm{diag}(\exp(\tilde{\mathbf{B}}\tilde{\boldsymbol{x}}_{ij}))\}^{-1}m^{-1}_{ij}\boldsymbol{\Gamma}_{ij}^T\bar{\boldsymbol{\Sigma}}^{-1/2}_{ij}
\end{align*}
which indicates that 
\[
m^{-1}_{ij}\boldsymbol{\Gamma}^T_{ij}\bar{\boldsymbol{\Sigma}}^{-1/2}_{ij}(\boldsymbol{y}_{ij}^{obs}-\boldsymbol{\Gamma}_{ij}\mathbf{A}\tilde{\boldsymbol{x}}_{ij}-\boldsymbol{\Gamma}_{ij}\boldsymbol{b}_i)\sim \mathcal{N}\left(\boldsymbol{0}, \mathrm{diag}(\exp(\tilde{\mathbf{B}}\tilde{\boldsymbol{x}}_{ij}))\right)
\]

Thus, the transformed representation converts inference on the latent covariance matrix into inference on parameters in tangent space, while retaining the contribution of all observed modalities.

The posterior distribution for parameters can be formulated as
\[
p(\boldsymbol{\Gamma},\mathbf{A},\mathbf{\tilde{B}},\boldsymbol{b},\boldsymbol{\Omega}_{\boldsymbol{b}}|\mathcal{D})\propto p(\boldsymbol{\Gamma},\mathbf{A},\mathbf{\tilde{B}},\boldsymbol{b},\boldsymbol{\Omega}_{\boldsymbol{b}})\prod_{i,j} p(\boldsymbol{y}_{ij}^{obs}|\boldsymbol{\Gamma},\mathbf{A},\mathbf{\tilde{B}},\boldsymbol{b},\boldsymbol{\Omega}_{\boldsymbol{b}},\tilde{\boldsymbol{x}}_{ij})
\] 
\[
\propto p(\mathbf{A})p(\mathbf{\tilde{B}})\prod_{i}p(\boldsymbol{b}_i|\boldsymbol{\Omega}_{\boldsymbol{b}})p(\boldsymbol{\Omega}_{\boldsymbol{b}})\prod_{k}p(\boldsymbol{\Gamma}^{(k)})\prod_{i,j} p(\boldsymbol{y}_{ij}^{obs}|\boldsymbol{\Gamma}_{ij},\mathbf{A},\mathbf{\tilde{B}},\boldsymbol{b}_i,\boldsymbol{\Omega}_{\boldsymbol{b}},\tilde{\boldsymbol{x}}_{ij})
\] 
where $\mathcal{D}$ denotes the observed data, and $\boldsymbol{b}$ denotes the collection of subject-specific random effects. For prior, $\boldsymbol{\Gamma}^{(k)}$ is individually sampled for each modality $k$ following matrix angular central Gaussian distribution uniform on the Stiefel manifold (\cite{jauch2021monte}), and weakly informative priors are applied for $\mathbf{A}$ and $\mathbf{\tilde{B}}$ that the prior for each element in the matrices follows $\mathcal{N}(0, 3)$. For random effects, $\boldsymbol{b}_i\sim \mathcal{N}(\boldsymbol{0}, \boldsymbol{\Omega}_{\boldsymbol{b}})$, where $\boldsymbol{\Omega}_{\boldsymbol{b}}$ is a diagonal matrix with the priors for all diagonal elements specified as $\mathcal{N}^+(0, 1)$. 

Posterior samples of the estimands (e.g., cross-covariance) are obtained by evaluating the target quantities at each iteration using the directly sampled parameters. For instance, let $\boldsymbol{\Sigma}_{kk'}(\tilde{\boldsymbol{x}}_{ij})=\boldsymbol{\Gamma}^{(k)}\boldsymbol{\Omega}(\tilde{\boldsymbol{x}}_{ij})\boldsymbol{\Gamma}^{(k')T}$. Then for posterior draw $r$, the corresponding cross-covariance draw is $\hat{\boldsymbol{\Sigma}}^{(r)}_{kk'}(\tilde{\boldsymbol{x}}_{ij})=
\hat{\boldsymbol{\Gamma}}^{(k;r)}\hat{\boldsymbol{\Omega}}^{(r)}(\tilde{\boldsymbol{x}}_{ij})\hat{\boldsymbol{\Gamma}}^{(k';r)T}$, and we report the point estimate $\hat{\boldsymbol{\Sigma}}_{kk'}(\tilde{\boldsymbol{x}}_{ij})$ as the median across the posterior samples (e.g., for $r = 1,\cdots, 2000$) for each element in the matrix. The element-wise credible interval can be reported for exploratory purposes to illustrate the strength of dependence. In addition to estimates conditional on a specific profile $\tilde{\boldsymbol{x}}$, marginal estimates can be obtained by integrating over selected covariates. For example, let $\tilde{\boldsymbol{x}}'$ denote the subset of covariates to be marginalized, and the marginal estimate can be expressed as $\int \hat{\boldsymbol{\Sigma}}_{kk'}(\tilde{\boldsymbol{x}})\,dF(\tilde{\boldsymbol{x}}')$, with the remaining covariates fixed at specified values.

\section{SIMULATION}\label{sec:simulation}
In this paper, we apply low-rank approximation to extract dominant signals from noise-contaminated data under challenging conditions, such as asynchronous multimodal responses and limited sample sizes. Accordingly, we conducted simulation studies to assess the model performance with respect to the following factors: (i) specified rank $d$ for model fitting; (ii) percentage of synchronous observations; and (iii) sample size. 
\subsection{Study setup}
\subsubsection{Data generation}
In each simulation run, we simulated $N$ subjects, where $N\in\{30,60\}$, and each subject had $5$ repeated measurements. We simulated bimodal outcome data $\boldsymbol{y}^{(1)}_{ij},\boldsymbol{y}^{(2)}_{ij}\in \mathbb{R}^{10}$, and defined the combined outcome vector $\boldsymbol{y}_{ij}=(\boldsymbol{y}^{(1)T}_{ij}, \boldsymbol{y}^{(2)T}_{ij})^T\in \mathbb{R}^{20}$. The design vector included an intercept and two covariates, $\tilde{\boldsymbol{x}}_{ij}=(1,x_{1i}, x_{2ij})^T$, with $x_{1i}\sim \mathrm{Bernoulli}(0.5)$ and $x_{2ij}\in \{0,0.25,0.5,0.75,1\}$ representing a binary subject-level group indicator and a scaled visit-time covariate, respectively.

Because real-world multimodal covariance matrices are generally full rank, we generated data from a full-rank covariance model and evaluated how well the proposed low-rank estimator recovers the dominant signals. Specifically, we randomly generated the full-rank eigenvector matrix $\boldsymbol{\Gamma}^*\in \mathbb{R}^{20\times 20}$ from the uniform distribution on the Stiefel manifold. Furthermore, each element in $\mathbf{A}\in \mathbb{R}^{20\times 3}$ and $\mathbf{B}\in \mathbb{R}^{20\times 3}$ was generated from Laplace$(0, 0.1)$. For random effects, $\boldsymbol{b}_i\sim \mathcal{N}(0, 0.1\mathbf{I}_{20})$. Together, in each simulation run, the pseudo-data were simulated by 
\begin{equation}\label{eq:simulation}\boldsymbol{y}_{ij}\sim \mathcal{N}\left(\boldsymbol{\Gamma}^*\mathbf{A}\tilde{\boldsymbol{x}}_{ij}+\boldsymbol{\Gamma}^*\boldsymbol{b}_i, \ \boldsymbol{\Gamma}^*\mathrm{diag}(\exp(\mathbf{B}\tilde{\boldsymbol{x}}_{ij})) \boldsymbol{\Gamma}^{*T}\right)\end{equation}

To emulate asynchronous observations, modality-specific missingness was introduced at random such that each observation may contain both modalities or only a single modality. Let $\boldsymbol{y}^{sim}_{ij}$ denote the resulting observation used for model fitting, and $\boldsymbol{y}^{sim}_{ij}\in \{\boldsymbol{y}_{ij}, \boldsymbol{y}^{(1)}_{ij},\boldsymbol{y}^{(2)}_{ij}\}$.
\subsubsection{Comparator}
We propose a multivariate method for estimating dependence structures that addresses practical challenges commonly encountered in the analysis of longitudinal multimodal data. Few off-the-shelf methods directly address this setting, especially when covariates are continuous or time-varying. When the covariates are categorical (e.g., sex, race), a common analytic strategy is to first fit mixed-effect models for each response variable and then calculate the pairwise residual covariance within each covariate-based stratum. Therefore, we set $x_1$ and $x_2$ to be categorical to compare against this empirical strategy.
 
\subsubsection{Evaluation metrics}
Let $\boldsymbol{\Sigma}(x_1, x_2)$ denote the covariance matrix in equation \eqref{eq:simulation}, with $\mathbf{R}(x_1, x_2)$ representing the corresponding correlation matrix, and let $\boldsymbol{\Theta}=\boldsymbol{\Gamma}^*\mathbf{A}$ denote the coefficient matrix for the fixed effects. In this work, multiple sets of parameters are estimated jointly, and the factors considered in the simulation study may affect them differently. Therefore, we evaluated the residual covariance blocks $\boldsymbol{\Sigma}_{[v, w]}(x_1, x_2)$ and correlation blocks $\mathbf{R}_{[v, w]}(x_1, x_2)$ separately, with $v=1,\cdots,10$ and $w=11,\cdots,20$ corresponding to the cross-modality off-diagonal block, and $v<w$ with $(v,w)\in \{1,\cdots,10\}^2 \cup \{11,\cdots,20\}^2$ corresponding to the within-modality diagonal blocks. The fixed-effect coefficients $\boldsymbol{\Theta}$ were also assessed. Random effects were introduced to account for individual heterogeneity and were not interpreted.

 For covariance and correlation blocks, we assessed the estimation loss across all ten covariate combinations $(x_1,x_2)$ formed by the two levels of $x_1$ and five grid values of $x_2$ as
\[\mathcal{L}(\hat{\boldsymbol{\Sigma}}_{[v,w]})=\frac1{10}\sum_{x_1,x_2}||\hat{\boldsymbol{\Sigma}}_{[v,w]}(x_1,x_2)-\boldsymbol{\Sigma}_{[v,w]}(x_1,x_2)||,\]
and correlation estimate $\hat{\mathbf{R}}_{[v,w]}$ was evaluated similarly. The fixed effect was examined by $||\hat{\boldsymbol{\Theta}}-\boldsymbol{\Theta}||$. We evaluated estimates from both the proposed method and the comparator approach, and for $||\cdot||$, both Frobenius and max norms were calculated.

\subsection{Simulation results}
\begin{figure}[htbp]
    \centering
    \includegraphics[width=1\textwidth]{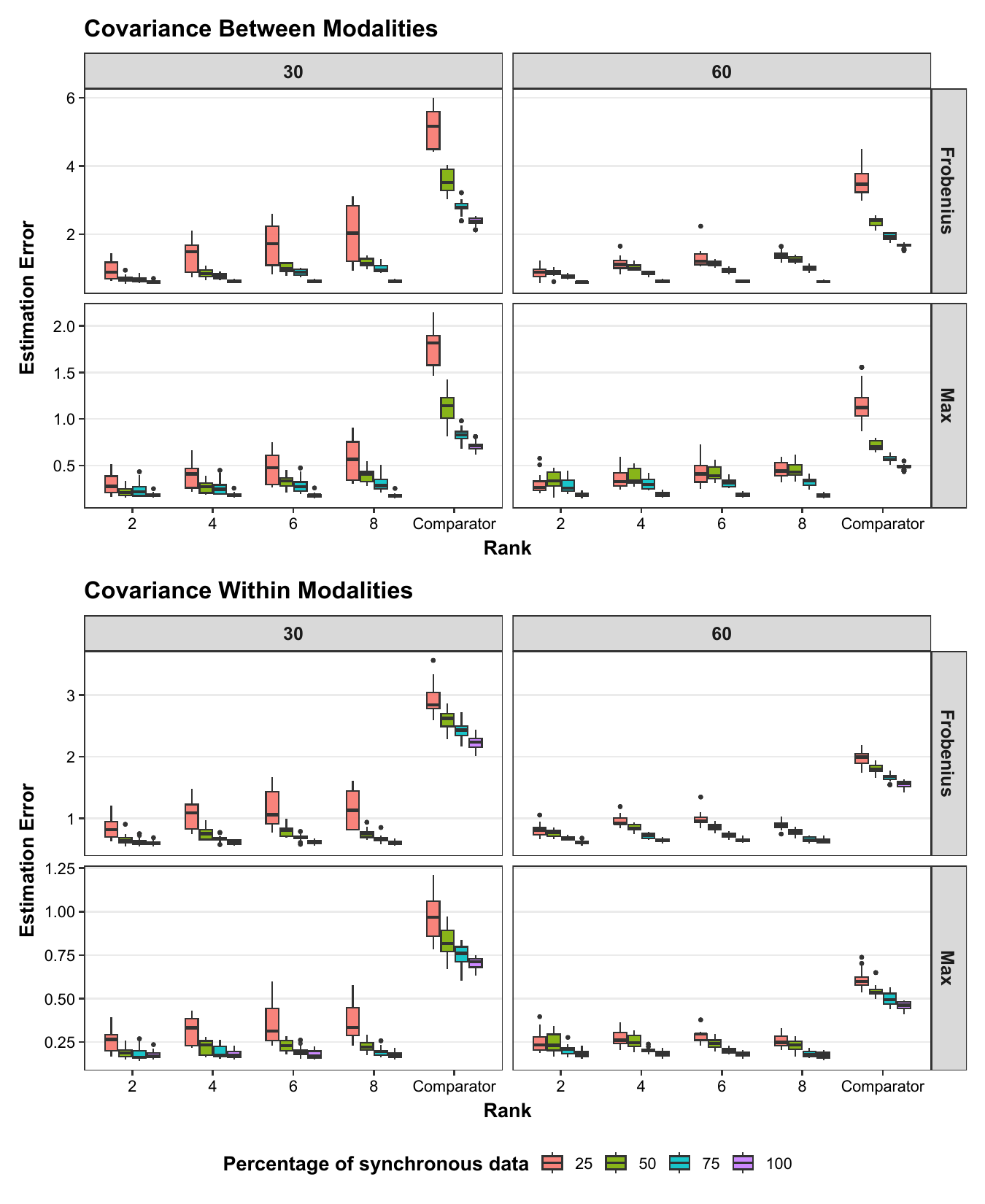}
    \caption{
        Estimation performance for covariance matrices, shown separately for the off-diagonal block representing cross-modality dependence (top panel) and the diagonal blocks representing within-modality dependence (bottom panel). The proposed method outperforms the comparator in covariance estimation, and the percentage of synchronous data has a greater influence than the other factors considered.
    }
    \label{fig:simulation-covariance}
\end{figure}

\Cref{fig:simulation-covariance,fig:simulation-correlation,fig:simulation-fixed} demonstrate the estimation performance for specified metrics across different ranks, percentages of synchronous observations $P(\boldsymbol{y}^{sim}_{ij}=\boldsymbol{y}_{ij})$, and sample sizes, where the boxplots summarize the distribution of estimation errors across 50 simulation runs. The summary statistics are provided in \Cref{tab:simulation-summary}. 
\subsubsection{Covariance}
As illustrated in \Cref{fig:simulation-covariance}, the proposed method consistently outperforms the comparator in covariance estimation across all simulation settings, regardless of the norm used in the evaluation. Specifically, within-modality diagonal blocks are estimated more accurately than the cross-modality off-diagonal block, and the effects of the percentage of synchronous observations, rank, and sample size are largely consistent across the estimation of both within- and cross-modality covariance. The estimation becomes more accurate as the percentage of synchronous observations increases. The rank for model fitting and sample size have mild overall impacts, while under settings with limited sample size and low synchrony, introducing a large rank may lead to unstable estimates, and a smaller rank may be preferable.

\subsubsection{Correlation}
\begin{figure}[htbp]
    \centering
    \includegraphics[width=1\textwidth]{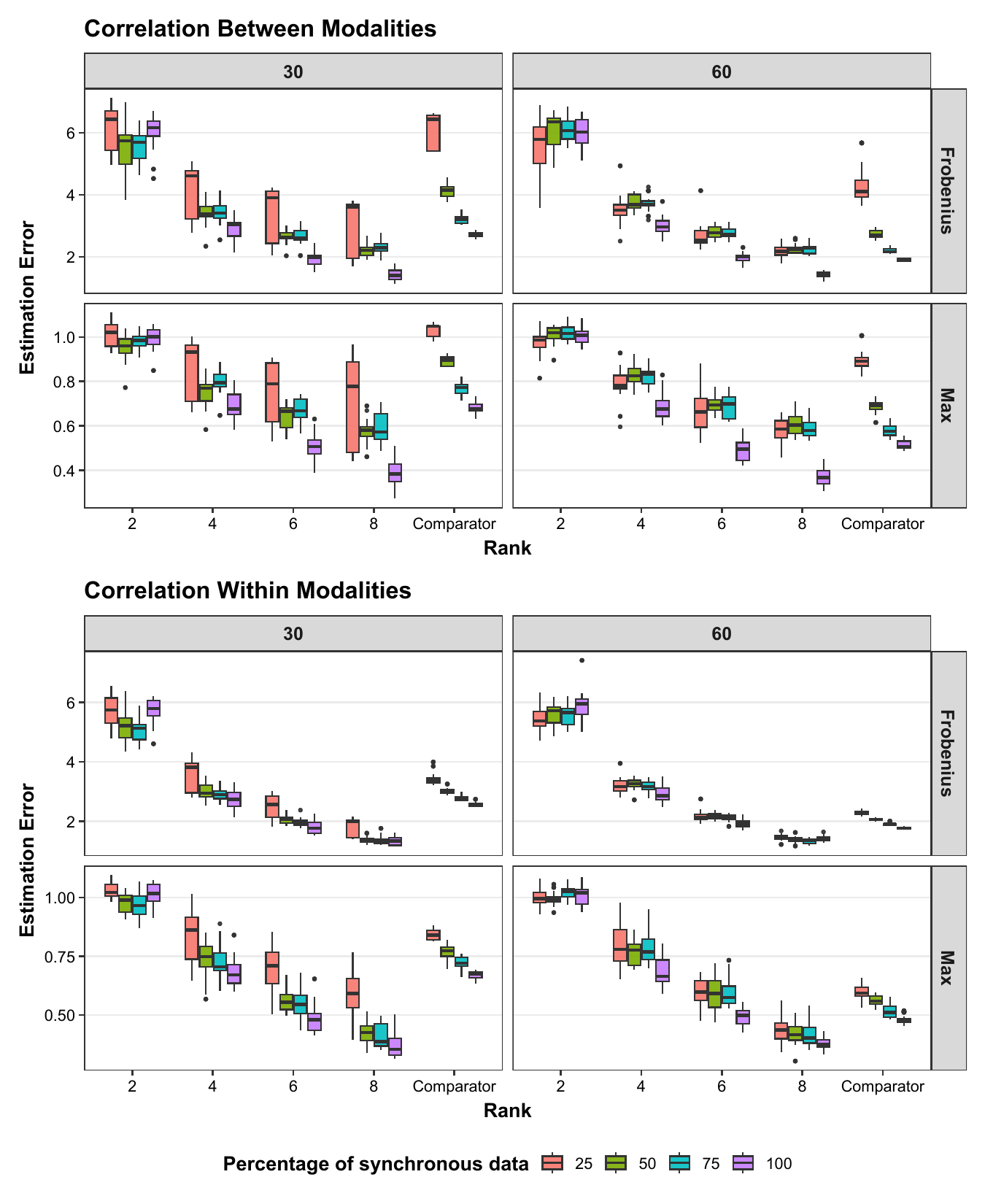}
    \caption{
            Estimation performance for correlation matrices, shown separately for the off-diagonal block representing cross-modality dependence (top panel) and the diagonal blocks representing within-modality dependence (bottom panel). Rank has a greater impact on correlation estimation, which generally benefits from higher-rank specifications.
    }
    \label{fig:simulation-correlation}
\end{figure}
\Cref{fig:simulation-correlation} suggests that a higher rank is preferable when accurate characterization of the correlation structure is of primary interest. By comparison, the impacts of sample size and percentage of synchronous observations are relatively mild. Although similar patterns are observed to those in covariance estimation that the use of higher ranks may result in unstable estimates under suboptimal conditions (e.g., sample size of 30 and 25$\%$ synchronous observations), especially for cross-modality correlation. Compared to the comparator, the proposed method remains effective in small sample settings, demonstrating superior performance even under relatively low-rank specifications.  
\subsubsection{Fixed effects}
\begin{figure}[htbp]
    \centering
    \includegraphics[width=1\textwidth]{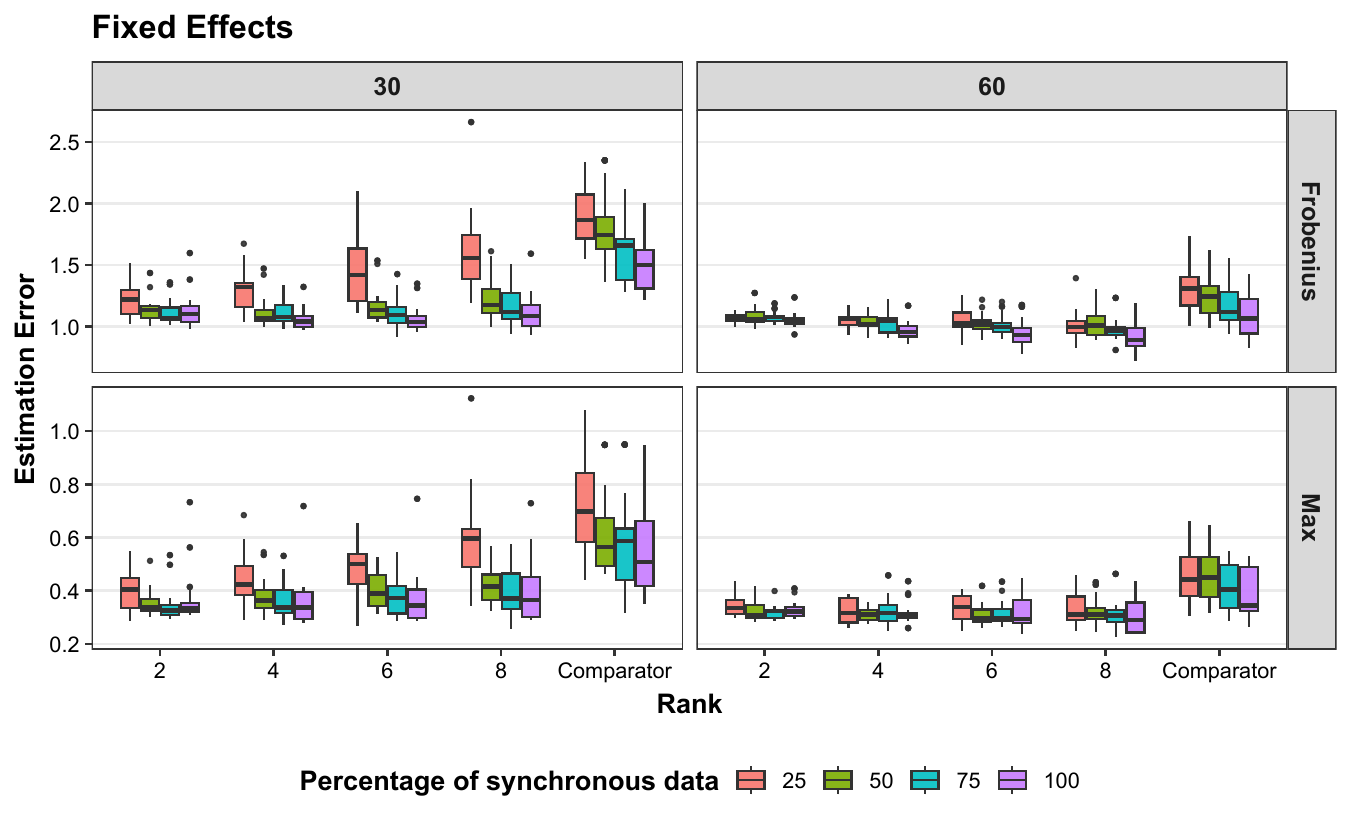}
    \caption{
        Estimation performance for fixed effects. Among factors evaluated in this simulation study, fixed effect estimation is more sensitive to sample size, with performance gains as the sample size increases.
    }
    \label{fig:simulation-fixed}
\end{figure}
The proposed model can yield more accurate and efficient estimates of fixed effects than standard regression models that assume homoskedasticity. As shown in \Cref{fig:simulation-fixed}, the estimation error from the proposed model is consistently lower than that from the comparator approach. Among the assessed factors, model performance improves as the sample size increases, with reduced bias and variability. In contrast, the percentage of synchronous data and the rank exhibit relatively modest effects. Consistent with the findings for covariance and correlation estimation, increasing the rank may not necessarily improve fixed effect estimation when the sample size is small, particularly under low levels of data synchrony. In such settings, a lower rank specification is generally adequate.
 
\subsubsection{Summary}
The relationship between the assessed factors (rank, percentage of synchronous observations, and sample size) and the estimation error of fixed effect, residual covariance, and correlation is inherently complex in the proposed model, because the likelihood is jointly maximized over multiple sets of parameters. In this simulation study, increasing the rank generally improves correlation estimation while having limited influence on fixed effect and covariance estimation. A larger sample size enhances fixed effect estimation, whereas its effects on covariance and correlation estimation are relatively modest. The percentage of synchronous observations moderately affects covariance and correlation estimation but has only a mild effect on fixed effect estimation.

Although real-world data may possess more complex underlying geometric structures, higher rank specifications are recommended when supported by sufficient data, as they may yield meaningful improvements in correlation estimation. Nevertheless, under limited-data conditions, large ranks can adversely affect model inference, in which case a lower rank specification is adequate and more stable.
\begin{table}[htbp]
    \centering
    \caption{
        Summary statistics of the simulation results. Entries report median (IQR) estimation loss over 50 simulation runs. Rows correspond to evaluation metrics and model ranks, and columns are grouped by sample sizes and degree of synchrony.
    }
    \includegraphics[width=0.92\textwidth]{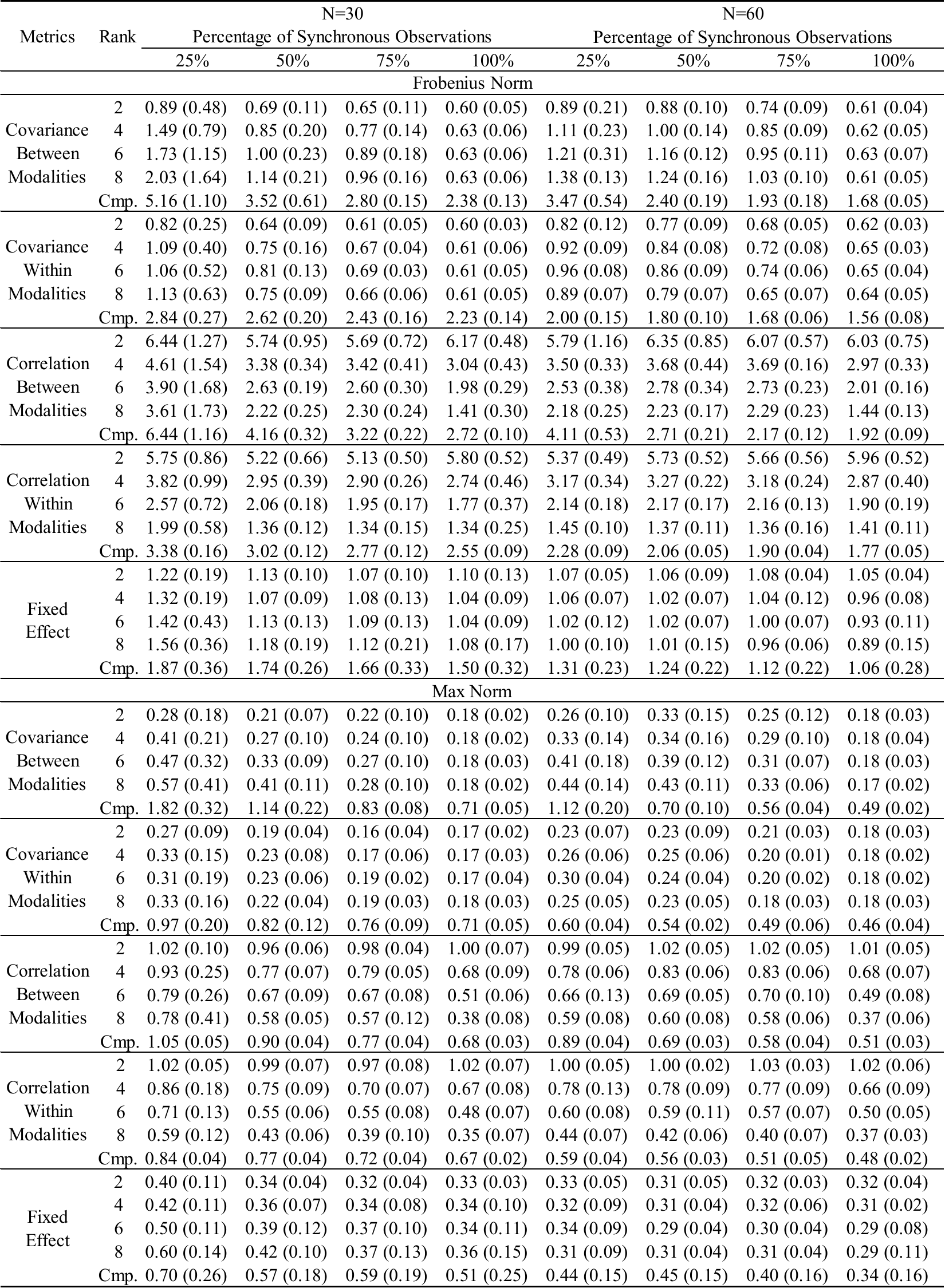}
    \label{tab:simulation-summary}
\begin{minipage}{0.95\linewidth}
\footnotesize
Note: \textit{Cmp.} in the Rank column indicates results from the comparator approach.
\end{minipage}
\end{table}

\section{APPLICATION}\label{sec:application}
We assessed the dependence across plasma and tau PET biomarkers in the ADNI data with the proposed model. Clinically, PET testing results are typically recommended for confirming disease status, but their use is limited by cost and accessibility (\cite{altomare2023plasma}). In comparison, plasma biomarkers are promising in the diagnosis and prognosis of AD because they are noninvasive and convenient to collect, while their performance in primary care requires further confirmation (\cite{hansson2022alzheimer}). In recent years, there has been growing interest in evaluating the pathological and clinical relevance of plasma biomarkers compared with imaging biomarkers (\cite{ossenkoppele2021tau}, \cite{ossenkoppele2022tau}, \cite{tissot2022comparing}, \cite{montoliu2025plasma}, \cite{karlsson2025machine}), and characterizing their temporal dynamics, longitudinal association, and timing of changes (\cite{rauchmann2021associations}, \cite{cogswell2024modeling}, \cite{mila2025timing}, \cite{yun2025temporal}).

Existing clinical evidence suggests that phosphorylated tau 217 (p-tau217) is strongly and positively associated with tau PET pathology (\cite{park2019plasma}, \cite{janelidze2021associations}, \cite{ashton2021validation}, \cite{pereira2021plasma}, \cite{meyer2022plasma}, \cite{mielke2022performance}, \cite{salvado2023specific}, \cite{altomare2023plasma}, \cite{cogswell2024modeling}, \cite{matthews2024relationships}, \cite{graff2025predictive}). The roles of amyloid beta (A$\beta$) 42 and A$\beta$40 have also been investigated extensively, and the ratio of A$\beta$42 to A$\beta$40 is typically more informative and generally negatively associated with tau PET (\cite{pereira2021plasma}, \cite{simren2021diagnostic}, \cite{ashton2021validation}, \cite{meyer2022plasma}, \cite{salvado2023specific}, \cite{graff2025predictive}, \cite{boutajangout2026association}). In addition, the ratio between p-tau and A$\beta$ has also been explored and is broadly positively associated with tau PET burden (\cite{park2019plasma}, \cite{boutajangout2026association}). Glial fibrillary acidic protein (GFAP) has been shown to exhibit positive association with tau PET uptake (\cite{chatterjee2022diagnostic}, \cite{simren2021diagnostic}, \cite{cogswell2024modeling}, \cite{matthews2024relationships}), and neurofilament light chain (NfL) levels may also increase with neurodegeneration progression, although their association with tau PET pathology is generally less specific (\cite{benedet2020stage}, \cite{ashton2021validation}, \cite{rauchmann2021associations}, \cite{chatterjee2022diagnostic}, \cite{matthews2024relationships}). 
\begin{figure}[htbp]
    \centering
    \includegraphics[width=0.7\textwidth]{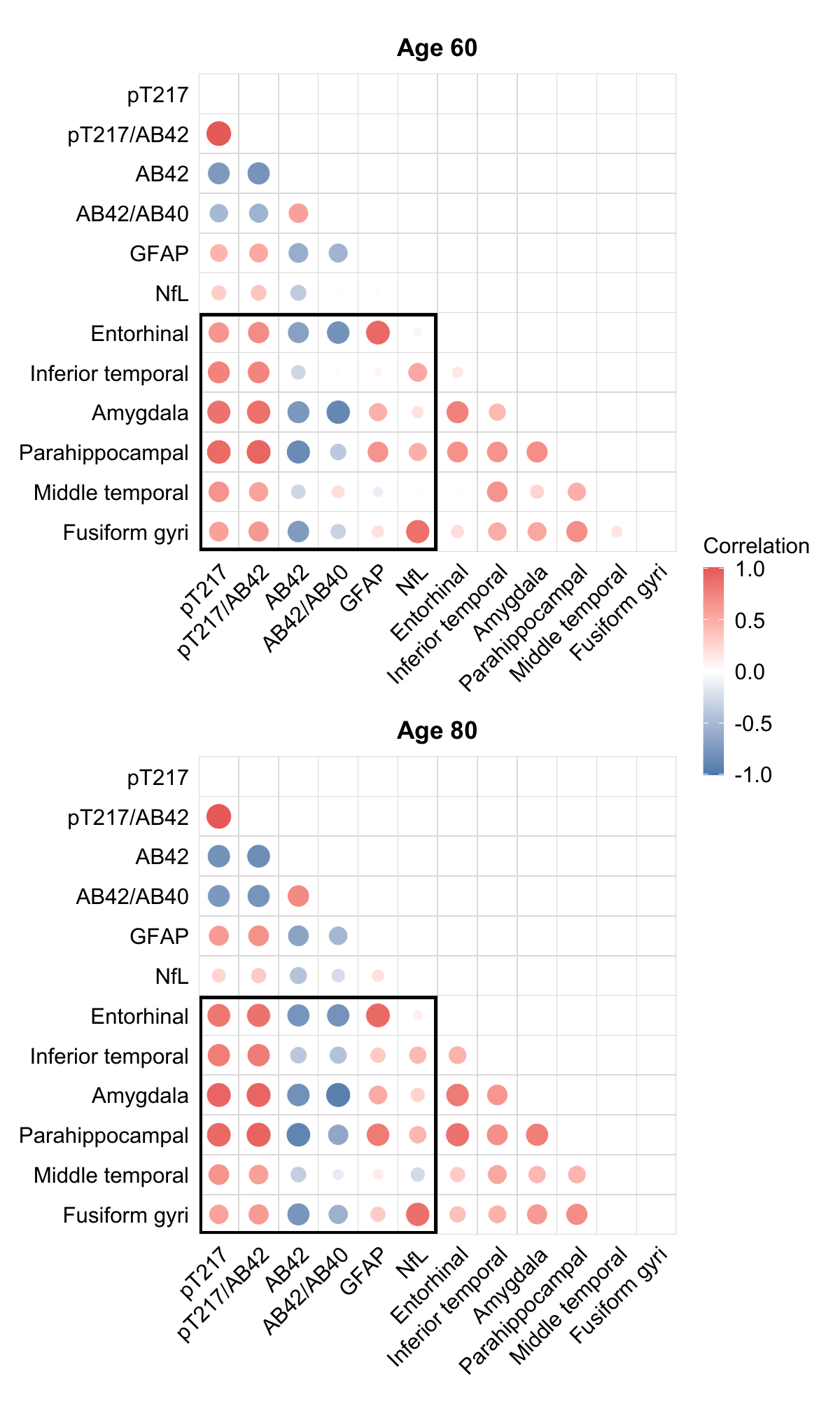}
    \caption{
        Estimated correlations among 6 plasma biomarkers and tau PET SUVR in 6 brain regions in the ADNI cohort at ages 60 (top panel) and 80 (bottom panel), adjusting for gender, APOE $\varepsilon$4 carrier status, race, and diagnostic status. The black box delineates the correlations across the two biomarker modalities.
    }
    \label{fig:application}
\end{figure}

Analytically, most clinical studies have focused on individual assessment of biomarkers or their joint utility for prediction, whereas the dependence structure across biomarkers remains less well understood. In this application study, we analyzed data obtained from the ADNI database (\url{adni.loni.usc.edu}). The ADNI was launched in 2003 as a public-private
partnership, led by Principal Investigator Michael W. Weiner, MD. The primary goal of ADNI has been to
test whether serial MRI, PET, other
biological markers, and clinical and neuropsychological assessment can be combined to measure the
progression of mild cognitive impairment (MCI) and early AD. Six plasma biomarkers (p-tau217, A$\beta$42, A$\beta$42/A$\beta$40, ptau217/A$\beta$42, NfL, and GFAP) that have been widely studied in AD were included, and tau PET standardized uptake value ratios (SUVRs) in six regions that are most often to be positive (entorhinal cortex, inferior temporal, the amygdalae, the parahippocampal gyri, the middle temporal, and the fusiform gyri) were incorporated with log transformation (\cite{st2024tau}). We applied the proposed multivariate method to a total of 490 longitudinal observations, where 125 ($25.5\%$) observations had both plasma and tau PET biomarker profiles, from 112 participants with either cognitively unimpaired or MCI diagnoses, and each participant contributed at least four repeated measurements. The visit-level covariates included diagnostic status (cognitively unimpaired and MCI) and age, and the subject-level covariates included gender, Apolipoprotein E (APOE) $\varepsilon4$ carrier status, and race.

\Cref{fig:application} displays the estimated correlation structure among the evaluated plasma and tau PET biomarkers at ages 60 and 80, and the observed trends align with previous studies. Specifically, for correlations across modalities, p-tau217 and p-tau217/A$\beta$42 ratio are positively associated with tau PET, whereas A$\beta$42 and A$\beta$42/A$\beta$40 are inversely associated with tau PET pathology. GFAP and NfL are broadly positively associated with tau PET. For correlations within the biomarker modalities, p-tau217, p-tau217/A$\beta$42, GFAP, and NfL tend to exhibit correlation patterns opposite to those of A$\beta$42 and A$\beta$42/A$\beta$40. In contrast, tau PET biomarkers are positively associated with one another across all evaluated brain regions. 
The age-specific results suggest that the associations among the assessed biomarkers generally become stronger with increasing age, while the overall dependence structure remains relatively stable at the population level. Standard stratified approaches cannot readily evaluate such dependence structures at specified values of continuous covariates. Furthermore, the proposed model provides the flexibility to characterize dependence structures for subgroups defined by covariates. In this application, however, the subgroup-specific differences were relatively modest, and we present results marginalized over the subgroup distributions for clarity. The present results are intended to illustrate the application of the proposed model, and a more comprehensive analysis of the ADNI data for clinical interpretation may require additional consideration of study design and data collection details, as well as covariate interaction effects. 

\section{DISCUSSION}\label{sec:discussion}
The proposed model characterizes covariate-specific dependence structures, measured through covariance or correlation matrices, among asynchronously observed random vectors in longitudinal studies. Several features of the proposed framework are worth emphasizing. First, the target of inference is a population-level covariance matrix indexed by time and demographic or clinical variables, rather than an individualized structure. Subject-specific random effects are incorporated to account for within-subject correlation and improve population-level inference. This setting is especially relevant for observational biomedical studies, where longitudinal measurements per subject are often not dense enough to support stable individualized estimation. In contrast, population-level estimation is generally more robust for characterizing overall dependence patterns. Second, the low-rank representation is intended to recover dominant shared dependence patterns across modalities, and therefore the method is well suited for real-world data with substantial noise. Third, because the likelihood is constructed from the observed modality blocks at each visit, the method can use partially observed multimodal records that would otherwise be discarded in complete-case analyses, thereby improving information use and estimation efficiency. Collectively, the proposed method provides a flexible framework for characterizing multimodal data from real-world longitudinal studies while accommodating common practical challenges, including limited longitudinal observations, noisy measurements, and asynchronous multimodal data collection. 

Although the proposed model is appropriate for settings where each subject is evaluated with a limited number of repeated measurements, convergence issues may arise and the estimation of random effects can become unstable when the observations are sparse with most subjects having fewer than three longitudinal measurements. In addition, the model performs well when the multimodal data signals are concentrated in a moderate number of dominant directions, which is the case in many real-world datasets. If the variance is distributed more evenly across many directions, posterior mixing tends to be slow, and estimation performance can be reduced. Further, higher rank specifications are generally recommended for real-world data applications because they can better capture complex underlying data structures, especially correlation patterns that are readily interpretable. However, in noisy or small-sample settings, higher-rank models may experience more frequent convergence issues. In such situations, lower rank specifications may provide a more stable alternative.

Further investigation is required to establish a hypothesis testing procedure to confirm the association between covariates and the dependence structure. In the current work, covariance matrices are modeled as covariate-dependent to depict dependence patterns at specified time points and subgroups rather than to conduct formal association tests. Hence, covariates primarily serve to index time points and subgroups for interpretation, and potential confounders can also be adjusted to facilitate exploratory comparisons by aligning confounder distributions. The development of a rigorous testing framework to identify key determinants of multimodal dependence could provide deeper insights into the underlying mechanisms. Moreover, the proposed model assumes Gaussian distributions for response vectors, which provides a reasonable working approximation for many continuous biomarkers, especially after appropriate transformation. However, it may be inadequate for outcomes with more complex distributional characteristics, such as bounded, zero-inflated, or discrete data. Further developments are needed to accommodate diverse forms of multimodal data encountered in practice.
\section*{SOFTWARE}
Software in the form of R codes and complete documentation are accessible at the following GitHub repository: \url{https://github.com/kqian2026/cmda}.
\section*{ACKNOWLEDGEMENTS}
Data collection and sharing for the Alzheimer's Disease Neuroimaging Initiative (ADNI) is funded by the National
Institute on Aging (National Institutes of Health Grant U19 AG024904). The grantee organization is the Northern
California Institute for Research and Education. In the past, ADNI has also received funding from the National
Institute of Biomedical Imaging and Bioengineering, the Canadian Institutes of Health Research, and private
sector contributions through the Foundation for the National Institutes of Health (FNIH) including generous
contributions from the following: AbbVie, Alzheimer’s Association; Alzheimer’s Drug Discovery Foundation;
Araclon Biotech; BioClinica, Inc.; Biogen; Bristol-Myers Squibb Company; CereSpir, Inc.; Cogstate; Eisai Inc.;
Elan Pharmaceuticals, Inc.; Eli Lilly and Company; EuroImmun; F. Hoffmann-La Roche Ltd and its affiliated
company Genentech, Inc.; Fujirebio; GE Healthcare; IXICO Ltd.; Janssen Alzheimer Immunotherapy Research \&
Development, LLC.; Johnson \& Johnson Pharmaceutical Research \&Development LLC.; Lumosity; Lundbeck;
Merck \& Co., Inc.; Meso Scale Diagnostics, LLC.; NeuroRx Research; Neurotrack Technologies; Novartis
Pharmaceuticals Corporation; Pfizer Inc.; Piramal Imaging; Servier; Takeda Pharmaceutical Company; and
Transition Therapeutics.
\section*{CONFLICT OF INTEREST} 
None declared.
\printbibliography[title={REFERENCES}]

@article{strimbu2010,
  title={What are biomarkers?},
  author={Strimbu, Kyle and Tavel, Jorge A},
  journal={Current Opinion in HIV and AIDS},
  volume={5},
  number={6},
  pages={463--466},
  year={2010},
  publisher={LWW}
}

@article{califf2018,
  title={Biomarker definitions and their applications},
  author={Califf, Robert M},
  journal={Experimental biology and medicine},
  volume={243},
  number={3},
  pages={213--221},
  year={2018},
  publisher={SAGE Publications Sage UK: London, England}
}

@article{schork1997,
  title={Genetics of complex disease: approaches, problems, and solutions},
  author={Schork, Nicholas J},
  journal={American journal of respiratory and critical care medicine},
  volume={156},
  number={4},
  pages={S103--S109},
  year={1997},
  publisher={American Thoracic Society New York, NY}
}

@article{johansson2023,
  title={Precision medicine in complex diseases—Molecular subgrouping for improved prediction and treatment stratification},
  author={Johansson, {\AA}sa and Andreassen, Ole A and Brunak, S{\o}ren and Franks, Paul W and Hedman, Harald and Loos, Ruth JF and Meder, Benjamin and Mel{\'e}n, Erik and Wheelock, Craig E and Jacobsson, Bo},
  journal={Journal of internal medicine},
  volume={294},
  number={4},
  pages={378--396},
  year={2023},
  publisher={Wiley Online Library}
}

@article{therriault2024,
  title={Biomarker-based staging of Alzheimer disease: rationale and clinical applications},
  author={Therriault, Joseph and Schindler, Suzanne E and Salvad{\'o}, Gemma and Pascoal, Tharick A and Benedet, Andr{\'e}a Lessa and Ashton, Nicholas J and Karikari, Thomas K and Apostolova, Liana and Murray, Melissa E and Verberk, Inge and others},
  journal={Nature Reviews Neurology},
  volume={20},
  number={4},
  pages={232--244},
  year={2024},
  publisher={Nature Publishing Group UK London}
}

@article{jack2024,
  title={Revised criteria for diagnosis and staging of Alzheimer's disease: Alzheimer's Association Workgroup},
  author={Jack, Jr., Clifford R. and Andrews, J. Scott and Beach, Thomas G. and Buracchio, Teresa and Dunn, Billy and Graf, Ana and Hansson, Oskar and Ho, Carole and Jagust, William and McDade, Eric and others},
  journal={Alzheimer's \& Dementia},
  volume={20},
  number={8},
  pages={5143--5169},
  year={2024},
  publisher={Wiley Online Library}
}

@article{leon2017,
  title={Probability, statistics, and random processes for electrical engineering},
  author={Leon-Garcia, Alberto},
  year={2017},
  publisher={Pearson Education}
}

@inproceedings{andrew2013,
  title={Deep canonical correlation analysis},
  author={Andrew, Galen and Arora, Raman and Bilmes, Jeff and Livescu, Karen},
  booktitle={International conference on machine learning},
  pages={1247--1255},
  year={2013},
  organization={PMLR}
}

@article{petersen2010,
  title={Alzheimer's disease Neuroimaging Initiative (ADNI) clinical characterization},
  author={Petersen, Ronald Carl and Aisen, Paul S and Beckett, Laurel A and Donohue, Michael C and Gamst, Anthony Collins and Harvey, Danielle J and Jack Jr, Clifford R and Jagust, William J and Shaw, Leslie M and Toga, Arthur W and others},
  journal={Neurology},
  volume={74},
  number={3},
  pages={201--209},
  year={2010},
  publisher={Lippincott Williams \& Wilkins}
}

@article{abe2022,
  title={Longitudinal structural brain changes in bipolar disorder: a multicenter neuroimaging study of 1232 individuals by the ENIGMA bipolar disorder working group},
  author={Ab{\'e}, Christoph and Ching, Christopher RK and Liberg, Benny and Lebedev, Alexander V and Agartz, Ingrid and Akudjedu, Theophilus N and Alda, Martin and Alnaes, Dag and Alonso-Lana, Silvia and Benedetti, Francesco and others},
  journal={Biological psychiatry},
  volume={91},
  number={6},
  pages={582--592},
  year={2022},
  publisher={Elsevier}
}

@article{edwards2023,
  title={Exploratory tau biomarker results from a multiple ascending-dose study of BIIB080 in Alzheimer disease: a randomized clinical trial},
  author={Edwards, Amanda L and Collins, Jessica A and Junge, Candice and Kordasiewicz, Holly and Mignon, Laurence and Wu, Shuang and Li, Yumeng and Lin, Lin and DuBois, Jonathan and Hutchison, R Matthew and others},
  journal={JAMA neurology},
  volume={80},
  number={12},
  pages={1344--1352},
  year={2023}
}

@article{he2025longitudinal,
  title={A longitudinal cohort study uncovers plasma protein biomarkers predating clinical onset and treatment response of rheumatoid arthritis},
  author={He, Siyu and Zhu, Chenxi and Liu, Yi and Xu, Zhiqiang and Sun, Rui and Yang, Bin and Guo, Xin and Herrmann i, Martin and Mu{\~n}oz, Luis E and Gjertsson, Inger and others},
  journal={Nature Communications},
  volume={16},
  number={1},
  pages={6692},
  year={2025},
  publisher={Nature Publishing Group UK London}
}

@article{trieu2025longitudinal,
  title={Longitudinal Blood-Based Biomarkers and Clinical Progression in Subjective Cognitive Decline},
  author={Trieu, Calvin and Van Harten, Argonde C and Van Leeuwenstijn, Mardou SSA and Schl{\"u}ter, Lisa-Marie and Boonkamp, Lynn and Aladdin, Azzam and Sikkes, Sietske AM and Van De Giessen, Elsmarieke and Verberk, Inge MW and Teunissen, Charlotte E and others},
  journal={JAMA Network Open},
  volume={8},
  number={12},
  pages={e2545862},
  year={2025},
  publisher={American Medical Association}
}

@article{veitch2024alzheimer,
  title={The Alzheimer's Disease Neuroimaging Initiative in the era of Alzheimer's disease treatment: a review of ADNI studies from 2021 to 2022},
  author={Veitch, Dallas P and Weiner, Michael W and Miller, Melanie and Aisen, Paul S and Ashford, Miriam A and Beckett, Laurel A and Green, Robert C and Harvey, Danielle and Jack Jr, Clifford R and Jagust, William and others},
  journal={Alzheimer's \& Dementia},
  volume={20},
  number={1},
  pages={652--694},
  year={2024},
  publisher={Wiley Online Library}
}

@article{weiner2025overview,
  title={Overview of Alzheimer's Disease Neuroimaging Initiative and future clinical trials},
  author={Weiner, Michael W and Kanoria, Shaveta and Miller, Melanie J and Aisen, Paul S and Beckett, Laurel A and Conti, Catherine and Diaz, Adam and Flenniken, Derek and Green, Robert C and Harvey, Danielle J and others},
  journal={Alzheimer's \& Dementia},
  volume={21},
  number={1},
  pages={e14321},
  year={2025},
  publisher={Wiley Online Library}
}

@article{kanoria2026clinical,
  title={Clinical impact of the Alzheimer's Disease Neuroimaging Initiative: A review of studies using ADNI data (2023 to June 2025)},
  author={Kanoria, Shaveta and Veitch, Dallas P and Miller, Melanie J and Aisen, Paul S and Beckett, Laurel A and Green, Robert C and Harvey, Danielle J and Jack Jr, Clifford R and Jagust, William and Lee, Edward B and others},
  journal={Alzheimer's \& Dementia},
  volume={22},
  number={4},
  pages={e71353},
  year={2026},
  publisher={Wiley Online Library}
}

@article{lloyd2019multi,
  title={Multi-omics of the gut microbial ecosystem in inflammatory bowel diseases},
  author={Lloyd-Price, Jason and Arze, Cesar and Ananthakrishnan, Ashwin N and Schirmer, Melanie and Avila-Pacheco, Julian and Poon, Tiffany W and Andrews, Elizabeth and Ajami, Nadim J and Bonham, Kevin S and Brislawn, Colin J and others},
  journal={Nature},
  volume={569},
  number={7758},
  pages={655--662},
  year={2019},
  publisher={Nature Publishing Group UK London}
}

@article{xiong2010binning,
  title={A binning method for analyzing mixed longitudinal data measured at distinct time points},
  author={Xiong, Xiaoqin and Dubin, Joel A},
  journal={Statistics in medicine},
  volume={29},
  number={18},
  pages={1919--1931},
  year={2010},
  publisher={Wiley Online Library}
}

@article{csenturk2013modeling,
  title={Modeling time-varying effects with generalized and unsynchronized longitudinal data},
  author={{\c{S}}ent{\"u}rk, Damla and Dalrymple, Lorien S and Mohammed, Sandra M and Kaysen, George A and Nguyen, Danh V},
  journal={Statistics in medicine},
  volume={32},
  number={17},
  pages={2971--2987},
  year={2013},
  publisher={Wiley Online Library}
}

@article{cao2015regression,
  title={Regression analysis of sparse asynchronous longitudinal data},
  author={Cao, Hongyuan and Zeng, Donglin and Fine, Jason P},
  journal={Journal of the Royal Statistical Society Series B: Statistical Methodology},
  volume={77},
  number={4},
  pages={755--776},
  year={2015},
  publisher={Oxford University Press}
}

@article{cao2016last,
  title={On last observation carried forward and asynchronous longitudinal regression analysis},
  author={Cao, Hongyuan and Li, Jialiang and Fine, Jason P},
  journal={Electronic Journal of Statistics},
  volume={10},
  pages={1155--1180},
  year={2016}
}

@article{chen2017analysis,
  title={Analysis of asynchronous longitudinal data with partially linear models},
  author={Chen, Li and Cao, Hongyuan},
  journal={Electronic Journal of Statistics},
  volume={11},
  pages={1549--1569},
  year={2017}
}

@article{sun2021regression,
  title={Regression analysis of asynchronous longitudinal data with informative observation processes},
  author={Sun, Dayu and Zhao, Hui and Sun, Jianguo},
  journal={Computational Statistics \& Data Analysis},
  volume={157},
  pages={107161},
  year={2021},
  publisher={Elsevier}
}

@article{li2023asynchronous,
  title={Asynchronous functional linear regression models for longitudinal data in reproducing kernel Hilbert space},
  author={Li, Ting and Zhu, Huichen and Li, Tengfei and Zhu, Hongtu},
  journal={Biometrics},
  volume={79},
  number={3},
  pages={1880--1895},
  year={2023},
  publisher={Wiley Online Library}
}

@article{li2022regression,
  title={Regression analysis of asynchronous longitudinal functional and scalar data},
  author={Li, Ting and Li, Tengfei and Zhu, Zhongyi and Zhu, Hongtu},
  journal={Journal of the American Statistical Association},
  volume={117},
  number={539},
  pages={1228--1242},
  year={2022},
  publisher={Taylor \& Francis}
}

@article{chang2023asynchronous,
  title={Asynchronous and error-prone longitudinal data analysis via functional calibration},
  author={Chang, Xinyue and Li, Yehua and Li, Yi},
  journal={Biometrics},
  volume={79},
  number={4},
  pages={3374--3387},
  year={2023},
  publisher={Wiley Online Library}
}

@article{liu2023regression,
  title={Regression analysis of mixed sparse synchronous and asynchronous longitudinal covariates with varying-coefficient models},
  author={Liu, Congmin and Sun, Zhuowei and Cao, Hongyuan},
  journal={Electronic Journal of Statistics},
  volume={17},
  number={2},
  pages={3103--3142},
  year={2023},
  publisher={The Institute of Mathematical Statistics and the Bernoulli Society}
}

@article{zhong2024locally,
  title={Locally sparse estimator of generalized varying coefficient model for asynchronous longitudinal data},
  author={Zhong, Rou and Zhang, Chunming and Zhang, Jingxiao},
  journal={Statistica Sinica},
  volume={34},
  number={4},
  pages={1903--1921},
  year={2024},
  publisher={JSTOR}
}

@article{lu2025semiparametric,
  title={Semiparametric mixture regression for asynchronous longitudinal data using multivariate functional principal component analysis},
  author={Lu, Ruihan and Li, Yehua and Yao, Weixin},
  journal={Biostatistics},
  volume={26},
  number={1},
  pages={kxaf008},
  year={2025},
  publisher={Oxford University Press}
}

@article{liu2026comprehensive,
  title={Comprehensive Analysis of Asynchronous Binary Variable Associations in Longitudinal End-of-Life Studies},
  author={Liu, Zhuangzhuang and Kim, Sanghee and Cho, Hyunkeun},
  journal={Statistics in Medicine},
  volume={45},
  number={3-5},
  pages={e70438},
  year={2026},
  publisher={Wiley Online Library}
}

@article{park2025bayesian,
  title={Bayesian estimation of covariate assisted principal regression for brain functional connectivity},
  author={Park, Hyung G},
  journal={Biostatistics},
  volume={26},
  number={1},
  pages={kxae023},
  year={2025},
  publisher={Oxford University Press}
}

@article{ng2015transport,
  title={Transport on Riemannian manifold for connectivity-based brain decoding},
  author={Ng, Bernard and Varoquaux, Gael and Poline, Jean Baptiste and Greicius, Michael and Thirion, Bertrand},
  journal={IEEE transactions on medical imaging},
  volume={35},
  number={1},
  pages={208--216},
  year={2015},
  publisher={IEEE}
}

@article{laird1982random,
  title={Random-effects models for longitudinal data},
  author={Laird, Nan M and Ware, James H},
  journal={Biometrics},
  pages={963--974},
  year={1982},
  publisher={JSTOR}
}

@article{backenroth2018modeling,
  title={Modeling motor learning using heteroscedastic functional principal components analysis},
  author={Backenroth, Daniel and Goldsmith, Jeff and Harran, Michelle D and Cortes, Juan C and Krakauer, John W and Kitago, Tomoko},
  journal={Journal of the American Statistical Association},
  volume={113},
  number={523},
  pages={1003--1015},
  year={2018},
  publisher={Taylor \& Francis}
}

@article{zhao2024longitudinal,
  title={Longitudinal regression of covariance matrix outcomes},
  author={Zhao, Yi and Caffo, Brian S and Luo, Xi},
  journal={Biostatistics},
  volume={25},
  number={2},
  pages={385--401},
  year={2024},
  publisher={Oxford University Press}
}

@article{jauch2021monte,
  title={Monte Carlo simulation on the Stiefel manifold via polar expansion},
  author={Jauch, Michael and Hoff, Peter D and Dunson, David B},
  journal={Journal of Computational and Graphical Statistics},
  volume={30},
  number={3},
  pages={622--631},
  year={2021},
  publisher={Taylor \& Francis}
}

@book{golub2013matrix,
  title={Matrix computations},
  author={Golub, Gene H and Van Loan, Charles F},
  year={2013},
  publisher={JHU press}
}

@article{altomare2023plasma,
  title={Plasma biomarkers for Alzheimer’s disease: a field-test in a memory clinic},
  author={Altomare, Daniele and Stampacchia, Sara and Ribaldi, Federica and Tomczyk, Szymon and Chevalier, Claire and Poulain, G{\'e}raldine and Asadi, Saina and Bancila, Bianca and Marizzoni, Moira and Martins, Marta and others},
  journal={Journal of Neurology, Neurosurgery \& Psychiatry},
  volume={94},
  number={6},
  pages={420--427},
  year={2023},
  publisher={BMJ Publishing Group Ltd}
}

@article{ashton2021validation,
  title={The validation status of blood biomarkers of amyloid and phospho-tau assessed with the 5-phase development framework for AD biomarkers},
  author={Ashton, NJ and Leuzy, A and Karikari, TK and Mattsson-Carlgren, N and Dodich, A and Boccardi, M and Corre, J and Drzezga, A and Nordberg, A and Ossenkoppele, R and others},
  journal={European journal of nuclear medicine and molecular imaging},
  volume={48},
  number={7},
  pages={2140--2156},
  year={2021},
  publisher={Springer}
}

@article{benedet2020stage,
  title={Stage-specific links between plasma neurofilament light and imaging biomarkers of Alzheimer’s disease},
  author={Benedet, Andrea L and Leuzy, Antoine and Pascoal, Tharick A and Ashton, Nicholas J and Mathotaarachchi, Sulantha and Savard, Melissa and Therriault, Joseph and Kang, Min Su and Chamoun, Mira and Sch{\"o}ll, Michael and others},
  journal={Brain},
  volume={143},
  number={12},
  pages={3793--3804},
  year={2020},
  publisher={Oxford University Press}
}

@article{boutajangout2026association,
  title={Association of plasma biomarkers with amyloid and tau PET in pre-dementia stages},
  author={Boutajangout, Allal and Masurkar, Arjun V and Osorio, Ricardo and Debure, Ludovic and Ghuman, Mobeena and Ahmed, Wajiha and Vedvyas, Alok and Pirraglia, Elizabeth and Links, Jon and Bokacheva, Louisa and others},
  journal={Alzheimer's \& Dementia},
  volume={22},
  number={5},
  pages={e71441},
  year={2026},
  publisher={Wiley Online Library}
}

@article{chatterjee2022diagnostic,
  title={Diagnostic and prognostic plasma biomarkers for preclinical Alzheimer's disease},
  author={Chatterjee, Pratishtha and Pedrini, Steve and Ashton, Nicholas J and Tegg, Michelle and Goozee, Kathryn and Singh, Abhay K and Karikari, Thomas K and Simr{\'e}n, Joel and Vanmechelen, Eugeen and Armstrong, Nicola J and others},
  journal={Alzheimer's \& Dementia},
  volume={18},
  number={6},
  pages={1141--1154},
  year={2022},
  publisher={Wiley Online Library}
}

@article{cogswell2024modeling,
  title={Modeling the temporal evolution of plasma p-tau in relation to amyloid beta and tau PET},
  author={Cogswell, Petrice M and Lundt, Emily S and Therneau, Terry M and Wiste, Heather J and Graff-Radford, Jonathan and Algeciras-Schimnich, Alicia and Lowe, Val J and Mielke, Michelle M and Schwarz, Christopher G and Senjem, Matthew L and others},
  journal={Alzheimer's \& Dementia},
  volume={20},
  number={2},
  pages={1225--1238},
  year={2024},
  publisher={Wiley Online Library}
}

@article{graff2025predictive,
  title={Predictive Value of Plasma Biomarkers in Tau-PET Transitions},
  author={Graff-Radford, Jonathan and Syrjanen, Jeremy A and Vemuri, Prashanthi and Lowe, Val J and Schwarz, Christopher G and Wiste, Heather J and Kremers, Walter K and Algeciras-Schimnich, Alicia and Pichet Binette, Alexa and Smith, Ruben and others},
  journal={Annals of neurology},
  volume={98},
  number={6},
  pages={1249--1260},
  year={2025},
  publisher={Wiley Online Library}
}

@article{hansson2022alzheimer,
  title={The Alzheimer's Association appropriate use recommendations for blood biomarkers in Alzheimer's disease},
  author={Hansson, Oskar and Edelmayer, Rebecca M and Boxer, Adam L and Carrillo, Maria C and Mielke, Michelle M and Rabinovici, Gil D and Salloway, Stephen and Sperling, Reisa and Zetterberg, Henrik and Teunissen, Charlotte E},
  journal={Alzheimer's \& Dementia},
  volume={18},
  number={12},
  pages={2669--2686},
  year={2022},
  publisher={Wiley Online Library}
}

@article{janelidze2021associations,
  title={Associations of plasma phospho-tau217 levels with tau positron emission tomography in early Alzheimer disease},
  author={Janelidze, Shorena and Berron, David and Smith, Ruben and Strandberg, Olof and Proctor, Nicholas K and Dage, Jeffrey L and Stomrud, Erik and Palmqvist, Sebastian and Mattsson-Carlgren, Niklas and Hansson, Oskar},
  journal={JAMA neurology},
  volume={78},
  number={2},
  pages={149--156},
  year={2021}
}

@article{karlsson2025machine,
  title={Machine learning prediction of tau-PET in Alzheimer's disease using plasma, MRI, and clinical data},
  author={Karlsson, Linda and Vogel, Jacob and Arvidsson, Ida and {\AA}str{\"o}m, Kalle and Strandberg, Olof and Seidlitz, Jakob and Bethlehem, Richard AI and Stomrud, Erik and Ossenkoppele, Rik and Ashton, Nicholas J and others},
  journal={Alzheimer's \& Dementia},
  volume={21},
  number={2},
  pages={e14600},
  year={2025},
  publisher={Wiley Online Library}
}

@article{matthews2024relationships,
  title={Relationships between plasma biomarkers, tau PET, FDG PET, and volumetric MRI in mild to moderate Alzheimer's disease patients},
  author={Matthews, Dawn C and Kinney, Jefferson W and Ritter, Aaron and Andrews, Randolph D and Toledano Strom, Erin N and Lukic, Ana S and Koenig, Lauren N and Revta, Carolyn and Fillit, Howard M and Zhong, Kate and others},
  journal={Alzheimer's \& Dementia: Translational Research \& Clinical Interventions},
  volume={10},
  number={3},
  pages={e12490},
  year={2024},
  publisher={Wiley Online Library}
}

@article{meyer2022plasma,
  title={Plasma p-tau231, p-tau181, PET biomarkers, and cognitive change in older adults},
  author={Meyer, Pierre-Fran{\c{c}}ois and Ashton, Nicholas J and Karikari, Thomas K and Strikwerda-Brown, Cherie and K{\"o}be, Theresa and Gonneaud, Julie and Pichet Binette, Alexa and Ozlen, Hazal and Yakoub, Yara and Simr{\'e}n, Joel and others},
  journal={Annals of neurology},
  volume={91},
  number={4},
  pages={548--560},
  year={2022},
  publisher={Wiley Online Library}
}

@article{mielke2022performance,
  title={Performance of plasma phosphorylated tau 181 and 217 in the community},
  author={Mielke, Michelle M and Dage, Jeffrey L and Frank, Ryan D and Algeciras-Schimnich, Alicia and Knopman, David S and Lowe, Val J and Bu, Guojun and Vemuri, Prashanthi and Graff-Radford, Jonathan and Jack Jr, Clifford R and others},
  journal={Nature medicine},
  volume={28},
  number={7},
  pages={1398--1405},
  year={2022},
  publisher={Nature Publishing Group US New York}
}

@article{mila2025timing,
  title={Timing of changes in Alzheimer's disease plasma biomarkers as assessed by amyloid and tau PET clocks},
  author={Mil{\`a}-Alom{\`a}, Marta and Tosun, Duygu and Schindler, Suzanne E and Hausle, Isabella and Petersen, Kellen K and Li, Yan and Dage, Jeffrey L and Du-Cuny, Lei and Saad, Ziad S and Saef, Benjamin and others},
  journal={Annals of neurology},
  volume={98},
  number={3},
  pages={508--523},
  year={2025},
  publisher={Wiley Online Library}
}

@article{montoliu2025plasma,
  title={Plasma tau biomarkers for biological staging of Alzheimer’s disease},
  author={Montoliu-Gaya, Laia and Salvad{\'o}, Gemma and Therriault, Joseph and Nilsson, Johanna and Janelidze, Shorena and Weiner, Sophia and Ashton, Nicholas J and Benedet, Andrea L and Rahmouni, Nesrine and Lantero-Rodriguez, Juan and others},
  journal={Nature Aging},
  volume={5},
  number={11},
  pages={2297--2308},
  year={2025},
  publisher={Nature Publishing Group US New York}
}

@article{ossenkoppele2021tau,
  title={Tau PET correlates with different Alzheimer’s disease-related features compared to CSF and plasma p-tau biomarkers},
  author={Ossenkoppele, Rik and Reimand, Juhan and Smith, Ruben and Leuzy, Antoine and Strandberg, Olof and Palmqvist, Sebastian and Stomrud, Erik and Zetterberg, Henrik and Alzheimer's Disease Neuroimaging Initiative and Scheltens, Philip and others},
  journal={EMBO molecular medicine},
  volume={13},
  number={8},
  pages={EMMM202114398},
  year={2021},
  publisher={Springer}
}

@article{ossenkoppele2022tau,
  title={Tau biomarkers in Alzheimer's disease: towards implementation in clinical practice and trials},
  author={Ossenkoppele, Rik and van der Kant, Rik and Hansson, Oskar},
  journal={The Lancet Neurology},
  volume={21},
  number={8},
  pages={726--734},
  year={2022},
  publisher={Elsevier}
}

@article{park2019plasma,
  title={Plasma tau/amyloid-$\beta$1--42 ratio predicts brain tau deposition and neurodegeneration in Alzheimer’s disease},
  author={Park, Jong-Chan and Han, Sun-Ho and Yi, Dahyun and Byun, Min Soo and Lee, Jun Ho and Jang, Sukjin and Ko, Kang and Jeon, So Yeon and Lee, Yun-Sang and Kim, Yu Kyeong and others},
  journal={Brain},
  volume={142},
  number={3},
  pages={771--786},
  year={2019},
  publisher={Oxford University Press}
}

@article{pereira2021plasma,
  title={Plasma markers predict changes in amyloid, tau, atrophy and cognition in non-demented subjects},
  author={Pereira, Joana B and Janelidze, Shorena and Stomrud, Erik and Palmqvist, Sebastian and Van Westen, Danielle and Dage, Jeffrey L and Mattsson-Carlgren, Niklas and Hansson, Oskar},
  journal={Brain},
  volume={144},
  number={9},
  pages={2826--2836},
  year={2021},
  publisher={Oxford University Press}
}

@article{rauchmann2021associations,
  title={Associations of longitudinal plasma p-tau181 and NfL with tau-PET, A$\beta$-PET and cognition},
  author={Rauchmann, Boris Stephan and Schneider-Axmann, Thomas and Perneczky, Robert},
  journal={Journal of Neurology, Neurosurgery \& Psychiatry},
  volume={92},
  number={12},
  pages={1289--1295},
  year={2021},
  publisher={BMJ Publishing Group Ltd}
}

@article{salvado2023specific,
  title={Specific associations between plasma biomarkers and postmortem amyloid plaque and tau tangle loads},
  author={Salvad{\'o}, Gemma and Ossenkoppele, Rik and Ashton, Nicholas J and Beach, Thomas G and Serrano, Geidy E and Reiman, Eric M and Zetterberg, Henrik and Mattsson-Carlgren, Niklas and Janelidze, Shorena and Blennow, Kaj and others},
  journal={EMBO molecular medicine},
  volume={15},
  number={5},
  pages={EMMM202217123},
  year={2023},
  publisher={Springer}
}

@article{simren2021diagnostic,
  title={The diagnostic and prognostic capabilities of plasma biomarkers in Alzheimer's disease},
  author={Simr{\'e}n, Joel and Leuzy, Antoine and Karikari, Thomas K and Hye, Abdul and Benedet, Andr{\'e}a Lessa and Lantero-Rodriguez, Juan and Mattsson-Carlgren, Niklas and Sch{\"o}ll, Michael and Mecocci, Patrizia and Vellas, Bruno and others},
  journal={Alzheimer's \& Dementia},
  volume={17},
  number={7},
  pages={1145--1156},
  year={2021},
  publisher={Wiley Online Library}
}

@article{st2024tau,
  title={Tau accumulation and its spatial progression across the Alzheimer’s disease spectrum},
  author={St-Onge, Fr{\'e}d{\'e}ric and Chapleau, Marianne and Breitner, John CS and Villeneuve, Sylvia and Pichet Binette, Alexa},
  journal={Brain Communications},
  volume={6},
  number={1},
  pages={fcae031},
  year={2024},
  publisher={Oxford University Press US}
}

@article{tissot2022comparing,
  title={Comparing tau status determined via plasma pTau181, pTau231 and [18F] MK6240 tau-PET},
  author={Tissot, Cecile and Therriault, Joseph and Kunach, Peter and Benedet, Andrea L and Pascoal, Tharick A and Ashton, Nicholas J and Karikari, Thomas K and Servaes, Stijn and Lussier, Firoza Z and Chamoun, Mira and others},
  journal={EBioMedicine},
  volume={76},
  year={2022},
  publisher={Elsevier}
}

@article{yun2025temporal,
  title={Temporal dynamics and biological variability of Alzheimer biomarkers},
  author={Yun, Jihwan and Shin, Daeun and Lee, Eun Hye and Kim, Jun Pyo and Ham, Hongki and Gu, Yuna and Chun, Min Young and Kang, Sung Hoon and Kim, Hee Jin and Na, Duk L and others},
  journal={JAMA neurology},
  volume={82},
  number={4},
  pages={384--396},
  year={2025}
}
\end{document}